\documentclass[preprint,3p]{elsarticle} 

\usepackage{amssymb}
\usepackage{amsthm}
\usepackage{amsmath}
\usepackage{parskip}

\usepackage{algorithm}
\usepackage[noend]{algorithmic}

\journal{Information Sciences}
\begin{document}
\begin{frontmatter}



\newtheorem{definition}{Definition}
\newtheorem{theorem}{Theorem}[section]
\newtheorem{lemma}[theorem]{Lemma}
\newtheorem{proposition}[theorem]{Proposition}
\newtheorem{corollary}[theorem]{Corollary}

\hyphenation{}

\title{Enhancing community detection using a network weighting strategy}


\author[1]{Pasquale De Meo \ }
\address[1]{University of Messina, Department of Physics, Informatics Section. V.le F. Stagno D'Alcontres 31, I-98166 Messina, Italy}
\ead{pdemeo@unime.it}
\author[2]{Emilio Ferrara \ \corref{cor1}}
\address[2]{Center for Complex Networks and Systems Research, School of Informatics and Computing.\\
Indiana University Bloomington, 919 E. 10th St., Bloomington, IN 47408, USA}
\ead{ferrarae@indiana.edu}
\cortext[cor1]{Corresponding author}
\author[1]{Giacomo Fiumara \ }
\ead{gfiumara@unime.it}
\author[1]{Alessandro Provetti \ }
\ead{ale@unime.it}

\address{}

\begin{abstract}
A community within a network is a group of vertices densely connected to each other but less connected
to the vertices outside.
The problem of detecting communities in large networks plays a key role in a wide range of research areas, e.g. Computer Science, Biology and Sociology.

Most of the existing algorithms to find communities count on the topological features of the network
and often do not scale well on large, real-life instances.

In this article we propose a strategy to enhance existing community detection algorithms by adding
a pre-processing step in which edges are weighted according to their \textsl{centrality} w.r.t. the network topology.
In our approach, the centrality of an edge reflects its contribute to making arbitrary graph tranversals, i.e., spreading messages over the network, as short as possible.
Our strategy is able to effectively complements information about network topology and it can be used as an additional tool to enhance community detection.
The computation of edge centralities is carried out by performing multiple random walks of bounded length on the network.
Our method makes the computation of edge centralities feasible also on large-scale networks.
It has been tested in conjunction with three state-of-the-art community detection algorithms, namely the Louvain method, COPRA and OSLOM.
Experimental results show that our method raises the accuracy of existing algorithms both on
synthetic and real-life datasets.
\end{abstract}

\begin{keyword}

Network Science \sep Complex Networks \sep Community Detection \sep Social Networks \sep Social Network Analysis
\end{keyword}

\end{frontmatter}

\section{Introduction}\label{sec:introduction}

Networks are a powerful tool to model real-life complex systems in many research fields like Biology, Sociology, Economy and Computer Science \cite{durugbo2011modelling,de2011recommendation}.
Due to their dynamics and sheer size, networks representing online social networks, e.g., Facebook, are a fascinating, and challenging, example of network models.

Most networks representing real-life systems show the so-called {\em community structure feature} \cite{fortunato2010community}: vertices tend to organize themselves in groups (called {\em communities} or {\em clusters}) such that the number of edges linking vertices of the same group is much higher than the number of edges joining vertices belonging to different groups.

The ability to detect a community within a larger network plays a key role in understanding how systems are organized: communities, in fact, can be regarded as {\em modules} whose functions or properties are, to some extent, separable from other modules.
The detection of communities is instrumental in understanding what are the main modules composing a real-life system, how these modules interact and, finally, how they evolve and impact the overall network and its functions.
Concrete examples from the biological domain rise from the task of understanding the functioning of metabolic networks \cite{GuAm05}, gene regulatory networks \cite{grunwald2008petri} or other forms of interactions among proteins \cite{von2003genome}.

In Computer Science and Sociology, community detection algorithms are a powerful tool to understand how humans interact.
There are, in fact, many reasons prompting users to join communities or to form new ones: people may decide to join a community because they share some interests with other community members \cite{chua2005personal} or because their attributes (like class or race) or cultural interests match well with those of the members of an already established community \cite{lin1986access}.
Finally, the Sociology literature shows that other factors (like ideologies, and attitudes of the members of a community) can prompt a user to join a community \cite{johnston2005social}.
Understanding the processes leading a user to join a community can, therefore, have a deep practical impact: for instance, in the design of advertisement and business applications, it is crucial to understand whether or not communities in a social network consist of persons sharing the same needs and tastes.
If such an hypothesis holds true it is possible to selectively disseminate commercial advertisements only to its members (who might be interested in those ads) rather than to the whole audience of the social network.

Due to its relevance, community detection has attracted the interest of many researchers and several, often interdisciplinary, approaches have been proposed.
Most of the existing community detection approaches aim at finding pairwise disjoint communities, i.e., communities that do not share members (represented by vertices of the network).
However, in the latest years, some researchers started studying the problem of finding {\em overlapping communities}, i.e., a relaxed version where a given vertex may belong to multiple communities \cite{Palla*05,gregory2007algorithm,lancichinetti2011finding}.

A major avenue to finding communities relies on the so-called {\em spectral clustering techniques} 	\cite{ng2001spectral,wei1989towards,hagen2002new}.
Spectral clustering aims at partitioning a graph into subsets of vertices, called {\em cuts.}
The problem of finding the optimal cuts is formulated as an optimization problem.
The main limitation of spectral clustering is that one has to know, or fix, in advance the number and the size of communities in the network.
Hence, this strategy is unfeasible when the purpose is to discover the unknown community structure of a network.

A further and relevant research line coincides with the introduction of a function called {\em network modularity} (usually denoted as $Q$) to quantitatively assess how structured in communities a given network is \cite{girvan2002community,clauset2004finding,duch2005community}.
In brief, the network modularity of a given network $\mathcal{N}$ is defined as the fraction of all edges that lie
within communities minus the expected value of the same quantity in a {\em random graph}
$\mathcal{N}^{'}$ so that: i) it has the same number of vertices of $\mathcal{N}$, ii) each vertex of $\mathcal{N}^{'}$ has the same degree of its peer in $\mathcal{N}$ and iii) edges are placed randomly with uniform probability.

The introduction of the network modularity allows to turn the problem of finding communities into an {\em optimization problem} whose goal is to find a partitioning of the network capable of maximizing $Q$.
Unfortunately, the maximization of $Q$ is an {\em NP-hard} problem \cite{brandes2007finding} thus  heuristics are required to find solutions (even near-optimal) and, at the same time, to guarantee reasonable computational costs/scalability.

Existing approaches based on modularity maximization suffer, however, from two major drawbacks.
The first drawback is that methods that are actually able to achieve high values for $Q$ work only on networks of small/medium size.
Consider, for instance, the first (and one of the most popular) algorithm to maximize modularity: the {\em Girvan-Newman} algorithm \cite{girvan2002community,newman2004finding}.
It iteratively removes edges, with the goal of partitioning the network into increasingly-disjoint communities.
Edges to be removed are selected according to their {\em betwenness centrality}: a measure of fraction of shortest paths between vertices that traverse that particular edge.
Which in turn is a computationally-heavy measure, as it depends directly on the number of vertices.
Hence, Girvan-Newman and similar methods are computationally expensive and do not scale well to the size of real-life networks consisting of, at least, millions of vertices and edges.
Indeed, methods explicitly designed to handle large networks are based on optimization techniques like {\em simulated annealing} \cite{guimera2004modularity} or {\em extremal optimization} \cite{duch2005community} and, therefore, the solution they produce may be sub-optimal.
The second drawback is the so-called {\em resolution limit} \cite{fortunato2007resolution}: communities consisting of a number of vertices smaller than a threshold (which in turn depends on the number of edges of the network) are not detected because the optimization procedure combines -- with the goal of maximizing \textit{Q} -- small groups of vertices into larger ones.

Several procedures have been proposed to alleviate the resolution limit problem such as providing novel definition of modularity \cite{li2008quantitative} or adding weights to the edges \cite{khadivi2011network,berry2011tolerating}.
To sum it up, despite the recent advances, community detection is still an open problem, even more so when we consider the growth, in sheer size and complexity, of online social networks.

In this article we propose a novel strategy to finding communities in networks which is based on the idea of introducing a measure of edge centrality and weighting edges according to their centrality.
Our ultimate goal, therefore, is meta-algorithmic: {\em not to introduce yet another community detection algorithm but to develop a pre-processing step devoted to weight the edges of the network.
Once the weights have been computed, existing community detection algorithms will execute with better results}.

The basis of our approach to the definition of edge weights, is the observation that in real-life social networks, a community can be intuitively depicted as a group of participants (vertices) which \textit{frequently interact with each other} (or at least more frequently than they do with third parties).
For instance, studies on online social networks like Facebook \cite{ferrara2011large,ferrara2012community} or Twitter \cite{grabowicz2012social} have shown that individuals belonging to the same community tend to frequently exchange messages with each other and seldom with people residing out of
the community itself.
This implies the existence of preferential pathways along which information flows easily.
Hence, social links can be ranked according to their capacity to facilitate the process of information propagation.

In fact, we believe that our approach represents a breakthrough because the methods considered until now {\em rely only} on the knowledge of the network topology whereas our approach suggests to complement information related to the network topology with information assessing the tendency of each edge to transfer information.

A parameter similar to our edge weighting was proposed by Fortunato et al. in \cite{FoLaMa04} and is called {\em efficiency}.
The efficiency of a pair of vertices \textit{i} and \textit{j} is defined as the inverse of the length of
the shortest path(s) connecting \textit{i} to \textit{j.}
The efficiency measure was used in the same paper to design a greedy algorithm to find communities; experimental results were carried out over real and artificial small networks to provide an evidence of the effectiveness of this parameter.

Unfortunately, efficiency can not be generalized to large-scale networks like Facebook.
In fact, to compute shortest paths, the whole network topology should be inspected and such an assumption
does not hold true in real social networks.
In any case, computing shortest paths in networks is costly, thus the computation of the network efficiency could be unfeasible over large networks.

Our approach is tailored to large networks and addresses those issues by means of the {\em random walks} technique to simulate message passing.
Random walks have been successfully exploited to simulate message passing in networks with the goal of computing {\em
node centrality} \cite{Newman2005} and, in this article, we propose to extend them to the computation of edge weights.

We execute multiple random walks and assign to edges a weight that is equal to the cumulative frequency of selecting that edge in a simulated random walk.
This choice is in keep with our previous considerations because, in our procedure, an edge has a high rank if it is frequently selected, i.e., if it is frequently exploited to convey messages.
In addition, paths generated in our simulations have to satisfy two further requirements: {\em (i)} an edge can be selected only once in a random walk in order to avoid that the weight of an edge may be excessively inflated, and {\em (ii)} the random walks consist of up to $\kappa$ edges,  $\kappa$ being a fixed integer.
In such a way, we acknowledge Friedkin's postulate \cite{friedkin1983horizons} that the more distant two vertices, the less they influence each other.
Moreover, the message propagation process is intended as a finite-steps process instead of a infinite one, which is reasonable in the context of real-life and online social networks where the spread of a given information sooner or later \textit{stops.}
The weight associated with each edge according to this strategy is called $\kappa$-{\em path edge centrality}.
Its accuracy as an estimate of an edge's \textit{importance} depends, of course, on several factors, including the number of random walks attempted and possible biases.
To the best of our knowledge, there are only two previous works concerning edge weighting in community detection, i.e., \cite{khadivi2011network} and \cite{berry2011tolerating}.
In those approaches the weight of a target edge is determined by computing some global network metrics like \textit{edge betwenness} \cite{khadivi2011network} or by considering cycles of fixed length \textit{k} that include the target edge \cite{berry2011tolerating}.
Inevitably, those approaches may not scale up well.
Our approach differs from those previous works since it exploits random walks to compute edge weights and, therefore, it does not require to know in advance the whole network topology.

To sum it up, these are the main contributions of this article:

\begin{itemize}

\item We provide a formal definition of $\kappa$-path edge centrality and describe an approximate algorithm called WERW-Kpath ({\em Weighted Edge Random Walk} - $\kappa$ {\em Path}) to efficiently compute the $\kappa$-path edge centrality.

\item We prove that our WERW-Kpath algorithm can approximate the actual value of centrality of an arbitrary edge with an error less than $\frac{1}{|E|}$,  $|E|$ being the number of edges
    in the network, by performing $O(\kappa|E|)$ iterations.
    The WERW-Kpath algorithm is fast because its computational complexity is \textsl{nearly-linear} in the number of edges of the network but, at the same time, it yields precise results.

\item We show how our edge weighting procedure can be combined with three existing, state-of-the-art algorithms for  community detection, namely the \textit{Louvain method} \cite{blondel2008fast,isda2011a}, COPRA \cite{gregory2007algorithm} and OSLOM \cite{lancichinetti2011finding}.

 \item We report on the experimental assessment of our approach on both real and artificial network datasets. In particular, we considered 9 real-world network datasets and the largest, a sample from Facebook, consists of 613,497 vertices and 2,045,030 edges.
Experiments on those networks show that combining WERW-Kpath with the algorithms mentioned above leads to an increase of network modularity up to 16\%.
			
Regarding the artificial networks, the experimental validation was conducted as follows: first, 72 artificial networks were generated, by exploiting the LFR benchmark \cite{lancichinetti2008benchmark}; in this way, we managed networks whose community structure was known in advance.
Then, we compared communities found by our approach in conjunction with the three methods above by using the so-called {\em Normalized Mutual Information} measure from Information Theory.
Experiments on these networks showed that our approach is able to alleviate the \textit{resolution limit} problem.
To make our research results reproducible, the prototype we implemented is freely available for download.

\end{itemize}

This article is organized as follows: in Section \ref{sec:background} we review existing approaches to finding communities in networks, whereas in Section \ref{sec:overview} we describe in detail our approach.
In Section \ref{sec:edgeweigthcomputation} we discuss the WERW-Kpath algorithm; Section \ref{sec:communitydetection} describes our proposal of adopting the WERW-KPath algorithm in conjunction with community detection algorithms to enhance their performance.
Section \ref{sec:experiments} is devoted to illustrate the experiments we carried out and to discuss the results.
Section \ref{sec:related} covers some related literature and, finally, in Section \ref{sec:conclusions}, we draw our conclusions and discuss possible developments.

\section{Background} \label{sec:background}
Recently, a huge amount of research work has concerned the detection of community structures inside networks.
In this section we describe some of the existing approaches to detecting communities; here and throughout this article we will use the terms {\em network} and {\em graph} interchangeably.
Of course, the material presented in this section cannot be exhaustive and we refer the reader to comprehensive surveys like \cite{porter2009communities,fortunato2010community}.

Given a network represented by a graph $G = \langle V, E \rangle$, the {\em community structure} is a partition $P = \{ C_1, C_2, \ldots, C_r \}$ of the vertices of $G$ such that, for each $C_i \in P$, the number of edges linking vertices in $C_i$ is high in comparison to the number of edges linking vertices on two distinct sets.
Each set $C_i$ is called {\em community}.

Today, the most popular techniques to find communities are: {\em (i) spectral clustering}, and, {\em (ii) network modularity maximization.}
In the following we shall discuss approaches belonging to each of these categories in detail.

\subsection{Spectral Clustering techniques}\label{sub:spectral}
Spectral Clustering techniques rely on the idea of partitioning a graph into subsets of vertices, called {\em cuts}.
The number of cuts to be generated is fixed in such a way as to minimize a given objective function.

The maximization of this objective function, however, has been proved to be {\em NP-hard.}
Therefore, different approximate techniques have been proposed.
For instance, in \cite{ng2001spectral}, the authors suggest to use the {\em Laplacian matrix} $L$ of a graph $G$.
We recall that the Laplacian matrix $L$ of $G = \langle V, E \rangle$ is a $|V| \times |V|$ matrix such that $L_{ij} = k_i\delta(i,j)-A_{ij}$, where $k_i$ is the degree of a vertex $i$, $\delta(i,j)$ is the {\em Kronecker symbol} (that is, $\delta(i,j)= 1$ if and only if $i=j$ and 0 otherwise) and $A_{ij}$ is the adjacency matrix of $G$.
Authors in \cite{ng2001spectral} propose to compute the top-$k$ eigenvectors of $L$, i.e., the eigenvector of $L$ associated with the $k$ eigenvalues having the largest magnitude.
The space of objects to cluster ({\em source space}) is then mapped onto the space generated by these eigenvectors ({\em target space}).
Finally, the \emph{k-means clustering algorithm} \cite{HaKa06} is applied on the points in the target space (with $k$ dimensions).

Another approach relies on the strategy of \textit{ratio cut partitioning} \cite{wei1989towards,hagen2002new}.
This is a function that, if minimized, allows the identification of large clusters with a minimum number of outgoing interconnections.
More recently, some authors \cite{mirkin2012additive} rely on the spectral decomposition of sparse matrices to
find communities.

The main issue with spectral clustering techniques is that one has to know in advance the number and the size of communities comprised in the given network.
This makes this strategy unfeasible if the purpose is to unveil the community structure of a network.
Finally, Shah et al. \cite{shah2010community} proved that this strategy could not work well if the given network
contains densely-connected yet small-sized communities.

\subsection{Network Modularity Maximization}\label{sub:networkmodularity}
Strategies based on network modularity define a measure, called {\em modularity} and usually denoted as \textit{Q,} to assess the quality of a partitioning of a graph $G$  and aim at finding the partition of \textit{G} that maximizes \textit{Q.}
Approaches based on network modularity rely on the idea that {\em random graphs} are not expected to exhibit a {\em community structure}.
Therefore, given a graph $G$ and a subgraph $C \subseteq G$, the {\em null model} $G^\prime$ associated with $G$ is defined as a graph having the same number of vertices of $G$ and obtained by preserving some of the structural properties of $G$.
For instance, $G^\prime$ could have the same number of edges of $G$ but these edges could be placed with a uniform probability among all pairs of nodes; in such a case $G^\prime$ is an example of a {\em Bernoulli random graph} \cite{fortunato2010community}.

Thanks to the {\em null model}, it is possible to decide whether a subgraph $C \subseteq G$ is a cluster or not.
In fact, since $G$ and $G^\prime$ have the same vertices, we can consider the subgraph $C^\prime \subseteq G^\prime$ obtained by isolating, in $G^\prime$, the vertices forming $C$ in $G$.
As claimed before, the null model is expected to \textit{exibit no community structure} and, therefore, we expect that $C^\prime$ is {\em not} a community.
Therefore, if the density of internal edges of $C$ is higher than that of $C^\prime$, we can conclude that $C$ is a community.

\noindent
Accordingly, the network modularity function is defined as follows

$$
Q = \frac{1}{2m} \sum_{i,j} \left( A_{ij} - P_{ij}\right) \delta(C_i,C_j).
$$

Here $m$ is the total number of edges in $G$, $A_{ij}$ is the adjacency matrix of $G$ (i.e., $A_{ij} = 1$ if there is an edge from $i$ to $j$ and 0 otherwise, $P_{ij}$ is the expected number of edges between $i$ and $j$ in the null model%
\footnote{Notice that $P_{ij}$ is a real number in $[0,1]$.}.
As usual, $\delta(\cdot,\cdot)$ is the Kronecker symbol.

Various null models are, in principle, allowed and, for each of them, we could derive a suitable expression for $P_{ij}$.
The most common choice, however, is to assume that $P_{ij}$ is proportional to the product of the degrees $k_i$ and $k_j$ of $i$ and $j$ respectively.
According to this choice, $Q$ can be rewritten as follows

\begin{equation}
\label{eqn:qmodexp}
Q = \frac{1}{2m}\sum_{i,j} \left(A_{ij} - \frac{k_i \cdot k_j}{2m}\right) \delta(C_i,C_j)
\end{equation}

Equation \ref{eqn:qmodexp} can be simplified by observing that only vertices belonging to the same community provide a non-zero contribution to the sum.
In fact, if $i$ and $j$ would belong to different communities then $C_i \neq C_j$ and $\delta(C_i,C_j) = 0$ by definition.
As a result, we can rewrite the modularity function $Q$ as follows

\begin{equation}	\label{eq:qmod}
	Q= \sum_{c = 1}^{n_c} \left[\frac{l_c}{m} - \left(\frac{d_c}{2m}\right)^2\right]
\end{equation}

where $n_c$ is the number of communities, $l_c$  is the total number of edges joining vertices
inside the community $c$ and $d_c$ is the sum of the degrees of the vertices composing $c$.
In Equation \eqref{eq:qmod}, for a fixed community $c$, the first term, i.e., $\frac{l_c}{m}$ (called {\em
coverage}) is the fraction of the edges of the graph inside $c$, whereas the second term $\left(\frac{d_c}{2m}\right)^2$ is the expected fraction of edges that would belong to $c$ in a random graph with the same degree distribution of $G$.

The problem of maximizing $Q$ has been proved to be {\em NP-hard} \cite{brandes2007finding}.
To this purpose, several heuristic strategies to maximize the network modularity $Q$ have been proposed as to date. Probably, the most popular one is known as the \emph{Girvan-Newman strategy} \cite{girvan2002community,newman2004finding}.

In this approach, edges are ranked by using a parameter known as {\em edge betweenness centrality}.
The edge betweenness centrality $B(e)$ of a given edge $e \in E$ is defined as

\begin{equation}	
	B(e) = \sum_{v_i \in V}\sum_{v_l \in V}\frac{np_e(v_i,v_l)}{np(v_i,v_l)}
	\label{eq:bc}
\end{equation}

where $v_i$ and $v_l$ are vertices in $V$, $np(v_i,v_l)$ is the number of shortest paths connecting $v_i$ and $v_l$ and $np_e(v_i, v_l)$ is the number of the shortest paths between $v_i$ and $v_l$ containing $e$.

Given the definition of edge betweenness centrality, it is possible to maximize the network modularity by {\em progressively deleting} edges with the highest value of betweenness centrality, based on the consideration that they shall connect vertices belonging to different communities \cite{newman2004finding}.
The process iterates until a significant increase of $Q$ is obtained.
At each iteration, each connected component of $G$ identifies a community.
Unfortunately, the computational cost of this strategy is $O(|V|^3)$ and this makes it unsuitable for the analysis of
large networks.
The most time-expensive part of the Girvan-Newman strategy is the calculation of the betweenness centrality.
Efficient algorithms have been designed to approximate the edge betweenness \cite{brandes2001faster}; for real-life networks the computational costs still remains prohibitive, unfortunately.

Several variants of this strategy have been proposed during the years, such as the \textit{Fast Clustering Algorithm} provided by Clauset, Newman and Moore \cite{clauset2004finding}, that runs in $O(|V| \log |V|)$ on sparse graphs.
In \cite{duch2005community}, Duch and Arenas proposed the {\em extremal optimization method} based on a fast agglomerative approach whose worst-case time complexity is $O(|V|^2 \log |V|)$.

An interesting network modularity maximization strategy is provided in the so-called \emph{Louvain method} (LM) \cite{blondel2008fast,isda2011a}.
LM has been tested in our experimental trials in conjunction with our approach to weighting edges; we present a detailed description of it in Section \ref{sec:communitydetection}.

The approaches mentioned above use {\em greedy strategies} to maximize $Q$.
In \cite{GuAm05} the authors propose to use {\em simulated annealing} to maximize $Q$.
This approach achieves a high accuracy but can as well be computationally very expensive.
In general terms, the advantage of simulated annealing techniques is that they do not suffer of the problem of getting stuck in local optima, differently from greedy algorithms.

\subsection{Finding Overlapping Communities}\label{sub:finding}
The algorithms presented above aim at finding {\em disjoint partitions}, i.e., partitions in which each vertex belongs to {\em exactly one} community.
It is interesting, also for practical purposes, to consider the relaxed case where vertices may happen to belong to different communities.
To clarify this concept, let us consider a network of researchers in Computer Science and observe that a researcher may belong to multiple communities like Database and Artificial Intelligence.
Communities sharing one or more vertices are said to be {\em overlapping} and the task of finding overlapping communities in networks has become one of the most popular research topics in Complex Networks research areas \cite{fortunato2010community}.

To the best of our knowledge, one of the first attempts to discover overlapping communities is due to Palla et al. \cite{Palla*05} who introduced {\em CFinder;} it detects communities by finding {\em cliques} of size $k$, where $k$ is a parameter provided by the user.
Such approach is time-expensive because the computational complexity of the clique detection is exponential in the number of involved nodes.
Experiments show that it scales well on real networks consisting up to $10^5$ nodes and, moreover, that it achieves a great accuracy.

Two other popular algorithms are COPRA ({\em Community Overlap PRopagation Algorithm}) \cite{gregory2007algorithm} and OSLOM ({\em Order Statistics Local Optimization Method}) \cite{lancichinetti2011finding}.
Both have been used in conjunction with our approach and, therefore, a detailed description of them will be presented in Section \ref{sec:communitydetection}.

\section{Overview of the Proposed Approach}\label{sec:overview}

In this section we discuss the technical and conceptual framework of our edge-weighting strategy.

We start by observing that in all approaches described in Section \ref{sec:background} consider
the network topology as the privileged (and, often, the only) source of information for community detection. Such an approach, however, runs contrary to the intuition  that, for example, in large online social networks like Twitter and Facebook communities should be identified as groups of users who frequently interact each other. Indeed, we expect that the volume of information exchanged among community members is
significantly higher than that exchanged between community members and people outside the community itself.

In contrast, the network topology only tells whether two users are connected, and therefore, if
two users are able to directly exchange messages or not; however, it {\em does not provide} any
indication whether two users actually communicate and, more in general, it does not inform us about
the existence of {\em preferential pathways} along which information flows. To better clarify this
concept, consider again networks like Facebook or Twitter. In both of them, a single user may have
a large number of contacts with whom she/he can exchange information (e.g., a \emph{wall post} on
Facebook or a \emph{tweet} on Twitter). Recent studies indicate that, in Facebook, the average
number of friends of a user is 120 but male users actively communicate with only 10 of them whereas women
with 16\footnote{{\tt http://www.economist.com/node/13176775?story\_id=13176775}}.

This implies that social links in a network vary in their ability to diffuse information over the network itself. Ranking is useful to
better understand how information flows among users.

We propose to {\em complement} information about network topology by weighting
each edge: the weight should indicate the ability of the edge itself to
transferring information. This supplementary source of knowledge will be later used (see Section
\ref{sec:communitydetection}) to find communities.

A concept similar to that introduced above has been already presented in Fortunato et al. \cite{FoLaMa04}. The authors assume that information from a vertex $i$ to a vertex $j$ travels along the
shortest paths connecting them. They define a parameter, called {\em efficiency}
$\varepsilon_{ij}$ as the inverse of the length of the shortest path connecting $i$ and $j$ and
used it to define a greedy algorithm to find communities. The algorithm identifies the edges that,
if removed, are able to generate the largest disruption in the network's ability to diffuse
information among its vertices and progressively delete them with the goal of splitting the network up
into communities.

In \cite{FoLaMa04}, the authors assume that information flows along shortest paths in networks. We
guess that such an assumption could not hold true in real scenarios: for instance, Facebook users are agnostic about the whole network topology and, practically, she/he
is not able to find shortest paths. In addition  the computation of information centrality requires
to calculate shortest paths and this activity can be prohibitively time-expensive in real networks.

To solve these drawbacks, we borrow some ideas that have been successfully applied in the past to
compute {\em node centrality}. In particular, Newman's suggestion to simulate message passing in
networks by means of {\em random walks}. A similar approach has been more recently considered in
\cite{alahakoon2011kpath}.

We simulate multiple random walks and assign the rank of an edge coincides with the
frequency of selecting the edge itself in all the simulations. This is compliant with our previous
reasoning since, in our procedure, an edge gets a high rank if it is frequently selected,
i.e., if it is frequently exploited to convey messages.

Our procedure has been designed to address the following requirements:

\paragraph{Simple Paths} We must avoid simulated random walks that pass more than once through an edge because this would
disproportionately inflate the rank of some edges while penalizing other ones.	

\paragraph{Bounded-Length Paths} As shown in \cite{friedkin1983horizons}, ``distant'' nodes in
    social networks (i.e., those nodes that are connected by long paths only) are unlikely
    to influence each other. We agree with this observation and figure that two nodes are
    considered to be distant if the path connecting them is longer than $\kappa$ hops,
    being $\kappa$ a fixed threshold. The impact of $\kappa$ on the performance of our approach
    will be extensively described in Section \ref{sec:experiments}.
    In addition, random walks of bounded length well represent the finiteness of the process of information propagation over large social networks.

The weight associated with each edge will be called $\kappa$-{\em path edge centrality}.
As a final comment, observe that our approach takes as input a graph $G = \langle V, E \rangle$ and
maps it onto a weighted graph $G^{'} = \langle V, E, W \rangle$ such that $W_{ij}$ is the weight of
the edge connecting vertices $i$ and $j$. Our approach is, therefore, {\em flexible} in the sense
that we can use it conjunction with any community detection algorithm. The community
detection algorithm will run on $G^{'}$ and, therefore, it can get a benefit from not only
information about network topology but also on information on edge centralities.


\section{Computation of edge weights} \label{sec:edgeweigthcomputation}

In this section we introduce an algorithm to assign weights to the edges of a network in compliance
with the requirements illustrated in Section \ref{sec:overview}. We first formalize the concept of
$\kappa$-path edge centrality (see Section \ref{sub:kpath-centrality}). After that, we describe an
approximate algorithm, called {\em WERW-Kpath} (see Section \ref{sub:WERWKpath}) for quickly
computing $\kappa$-path edge centralities. Finally, in Section \ref{sub:formalanalysis} we formally
analyze the accuracy achieved by the WERW-Kpath algorithm.

\subsection{The {\large $\kappa$}-Path Edge Centrality}
\label{sub:kpath-centrality}

The concept of {\em $\kappa$-path edge centrality} extends the concept of {\em $\kappa$-path vertex
centrality} introduced  for the first time in \cite{alahakoon2011kpath}. The $\kappa$-path vertex
centrality relies on the concept of {\em simple $\kappa$-path}:

\begin{definition} (simple $\kappa$-path). {\em Let $G = \langle V, E \rangle$ be a graph and
let $\kappa > 0$ be an integer. A simple $\kappa$-path is a simple path comprising at most $\kappa$
edges in $G$ and these edges are selected at random.} \label{def:simple-k-path}
\end{definition}

We are now able to formally introduce the notion of $\kappa$-path vertex centrality
\cite{alahakoon2011kpath}.

\begin{definition} ($\kappa$-path vertex centrality). {\em Let: {\em (i)} $G = \langle V, E \rangle$ be a graph,
{\em (ii)} $\kappa >0$ be an integer and {\em (iii)} $v_n \in V$ be a vertex in $G$. The
$\kappa$-path vertex centrality $C^\kappa(v_n)$ of $v_n$ is the sum, over all possible source
vertices $s$, of the probability with which a message originated from $s$ goes through $v_n$,
assuming that the message traversals are only along simple $\kappa$-paths.}
\label{def:k-path-centrality}
\end{definition}

The concept of $\kappa$-path vertex centrality can be extended so that to assess the
relevance of an edge in a network. Such a concept is formalized in Definition
\ref{def:kpathedgecentrality}.

\begin{definition} ($\kappa$-path edge centrality). {\em Let: {\em (i)} $G = \langle V, E \rangle$ be a graph,
{\em (ii)}  $\kappa >0$ be an integer and {\em (iii)} $e \in E$ be an edge of $G$. The
$\kappa$-path edge centrality $L^\kappa(e)$ of $e$ is the sum, over all possible source vertex $s$,
of the probability with which a message originated from $s$ traverses $e$, assuming that the
message traversals are only along random simple $\kappa$-paths.}\label{def:kpathedgecentrality}
\end{definition}

In the following, we shall introduce an algorithm, called {\em WERW-Kpath} to efficiently
compute $\kappa$-path edge centrality.

\subsection{The WERW-Kpath algorithm}
\label{sub:WERWKpath}

In this section we present the {\em WERW-Kpath} ({\em Weighted Edge Random Walk} - $\kappa$ {\em
Path}), an approximate algorithm to efficiently compute the edge centrality.

The WERW-Kpath algorithm takes a graph $G = \langle V, E\rangle$ and an integer $\kappa$ as input
and simulates $\rho$ {\em simple random paths} of at most $\kappa$ edges on $G$ such that the
length of each random walk is no greater than $\kappa$. Here $\rho$ is a fixed integer whose tuning
will be discussed later. At the beginning, a weight $\omega(e) = 1$ is assigned to each edge $e \in
E$. In each simulation, a source node $s$ is selected and $s$ is assumed to inject a message in
$G$; after that, $s$ selects, according to some strategy, one of its neighboring vertices, say $w$,
and forwards it the message. The weight of the edge connecting $s$ and $w$ is increased by 1 and
the process restarts from $w$.

Owing to this informal discussion, it emerges that the key ingredients of the WERW-Kpath algorithm
are the strategies exploited to select the starting node and the edges invoked to convey the
message. These strategies define a {\em message propagation model} on $G$. In the WERW-Kpath
algorithm we consider a model relying on two main assumptions.

The first assumption is that vertices are not all {\em equally relevant} to generate and spread
messages. In detail, we decide to privilege vertices representing users with a high level of
engagement in the social networks because we assume that the higher the number of connections of a
user, the more marked her/his aptitude to generate and spread messages. To quantitatively encode
this intuition we defined, for each vertex $v_n \in V$, the \emph{normalized degree} $\delta(v_n)$
of $v_n$ as follows

$$
\delta(v_n) = \frac{|I(v_n)|}{|V|}
$$

being $I(v_n)$ the set of edges incident onto $v_n$.

The normalized degree $\delta(v_n)$ correlates the degree of $v_n$ to the total number of vertices
in the network. It ranges in the real interval $[0,1]$ and the higher $\delta(v_n)$, the better
$v_n$ is connected in the graph. We suggest that the probability of selecting $v_n$ as the source
vertex is proportional to $\delta(v_n)$.

The second assumption is that edges are not all {\em equally relevant} in conveying messages. This
is compliant with the fact that, in real life, a user does not select {\em totally at random} the
user to whom a message should be forwarded but she/he selects the addressee according to some
criteria. These criteria may be formulated by assuming that edges in $G$ are weighted and requiring
that a user selects an edge to convey a message on the basis of the edge weights. Unfortunately,
weights on edges should be the {\em output} of the algorithm and not its {\em input} and, then, we do not
have at our disposal weights allowing us to select edges.

We solved this issue by observing that, in the WERW-KPath algorithm, edge weights are initially uniformly
assigned because all edges are deemed equally important in transferring
messages; after that, the weight of an edges will be increased if that edge is selected to convey a
message.

If we would stop the WERW-Kpath algorithm after $\ell$ iterations (with $\ell < \rho$), the weight
currently associated with that edge would represent an approximation of its actual weight. Of
course, the higher $\ell$, the better this approximation.

The weight assigned to each edge at the arbitrary, $\ell$-th iteration, is therefore an
approximation of its centrality value obtained by stopping the simulation procedure after $\ell$-th
iterations. Due to the definition of edge centrality, the WERW-Kpath algorithm has to select, in
each iteration, the edge having the highest centrality value and, therefore, the highest weight.

More formally, in the $\ell$-th iteration, we propose that the probability $\Pr(e_m)$ of selecting
an edge $e_m$ is proportional to its current edge weight $\omega(e_m)$

\begin{equation}
	\displaystyle{\Pr(e_m)=\frac{\omega(e_m)}{\sum_{e_m \in \hat{I}(v_n)} \omega(e_m)}}
	\label{eq:p-e-n}
\end{equation}

being $\hat{I}(v_n) = \{ e_k \in I(v_n) \ | \ T(e_k)=0 \}$ where $T(e_k)$ is defined as follows

\begin{equation}
\label{eqn:flag}
T(e_m) = \left\{
			\begin{aligned}
			&1 \quad  \mbox{if $e_m$  has   already  been  traversed}\\
			&0 \quad \mbox{otherwise.}
			\end{aligned}\right.
\end{equation}

In this way, the weight already awarded to an edge $e$ is assumed to be an indicator of its
tendency to transfer messages.

By putting together all these ideas we are able to provide a more formal description of the
WERW-Kpath algorithm. It takes a graph $G = \langle V, E\rangle$ as input and, as previously
pointed out, it assigns each edge $e_m \in E$ with a weight $\omega(e_m) = 1$.

After that, it iterates the following sub-steps a number of times equal to $\rho$, being $\rho$ a
fixed value (we will study in Section \ref{sub:formalanalysis} the impact of $\rho$ on the
performance of the algorithm):

\begin{enumerate}
		
\item A vertex $v_n \in V$ is selected at random with a probability $\Pr(v_n)$ proportional to
    $\delta(v_n)$.

\item  All the edges in $E$ are marked as {\em not traversed}.

\item The procedure {\em MessagePropagation} is invoked. It generates a simple random walk
    starting from $v_n$ whose length is not greater than $\kappa$.

\end{enumerate}

Let us describe the procedure {\em MessagePropagation}. For the purpose of readability, the
pseudo-code of the {\em MessagePropagation} procedure is reported in Algorithm \ref{alg:procedure}.

\begin{algorithm}
	\caption{MessagePropagation($v_n$: a Vertex , $N$: an integer, $\kappa$: an integer,
$\mathbf{\omega}$: an array of weights )}
	\label{alg:procedure}
	\begin{algorithmic}[1]
				\WHILE{$N < \kappa$ and $\left[|I(v_n)| > \sum_{e_k \in I(v_n)}{T(e_k)}\right]$}
                \STATE $e_{m} \leftarrow$ $e_m \in \{I(v_n)\ | \ T(e_m) = 0\}$,
                chosen with probability  given by Equation \eqref{eq:p-e-n}.
				\STATE Let $v_{n+1}$ be the vertex reached by $v_n$ through $e_m$
                \STATE $\omega(e_{m}) \leftarrow \omega(e_{m}) + 1$
				\STATE $T(e_{m}) \leftarrow  1$
				\STATE $v_{n} \leftarrow  v_{n+1}$
				\STATE $N \leftarrow N+1$
			\ENDWHILE
\end{algorithmic}
\end{algorithm}

This procedure carries out a loop until {\em both} the following conditions hold true:

\begin{enumerate}

\item The length of the path currently generated is no greater than $\kappa$.
			This is managed through a length counter $N$.

\item Assuming that the walk has reached the vertex $v_n$, the loop continues if there exists
    {\em at least one edge} incident on $v_n$ which has not been already traversed.

Since $I(v_n)$ is the set of edges incident on $v_n$, the following condition must be true:

\begin{equation}
	|I(v_n)| > \sum_{e_k \in I(v_n)}{T(e_k)}
	\label{eq:stop}
\end{equation}

\end{enumerate}

The former condition allows us to consider only paths up to length $\kappa$.
The latter avoids that the message passes more than once through an edge.

If the conditions above are satisfied, the {\em MessagePropagation} procedure selects an edge $e_m$
at random, with a probability $\Pr(e_m)$ given by Equation \eqref{eq:p-e-n}.

Let $e_m$ be the selected edge and let $v_{n+1}$ be the vertex reached from $v_n$ by means of
$e_m$. The {\em MessagePropagation} procedure increases $\omega(e_m)$ by 1, sets $T(e_m) = 1$ and
increases the counter $N$ by 1. The message propagation re-starts from $v_{n+1}$.

At the end, for each edge $e_m \in E$, the WERW-Kpath algorithm sets $\hat{L}^\kappa(e) \leftarrow
\frac{\omega(e_m)}{\rho}$. This value will be adopted as 	the {\em centrality} of $e_m$.

The pseudocode describing the WERW-Kpath algorithm is reported in Algorithm \ref{alg:ERW-Kpath}.

\begin{algorithm}
	\caption{WERW-Kpath($G=  \langle V,E \rangle$: a Graph, $\kappa$: an integer, $\rho$: an
integer)}
	\label{alg:ERW-Kpath}
	\begin{algorithmic}[1]
		\STATE For each $e_m \in E$ set $\omega(e_m) \leftarrow 1$
		\FOR{$i=1$ to $\rho$}			
			\STATE $N \leftarrow 0$ a counter to check the length of the $\kappa$-path
			\STATE $v_n \leftarrow$ a node chosen uniformly at random
			\STATE MessagePropagation($v_n$, $N$, $\kappa$, $\mathbf{\omega}$)
		\ENDFOR
        \STATE For each $e_m \in E$ set $\hat{L}^\kappa(e_m) \leftarrow
\frac{\omega(e_m)}{\rho}$
	\end{algorithmic}
\end{algorithm}

\subsection{Formal Analysis}
\label{sub:formalanalysis}

In this section we investigate to what extent the value $\hat{L}^\kappa(e)$ returned by the
WERW-Kpath algorithm is a ``good'' approximation of the actual centrality value of an edge provided
in Definition \ref{def:kpathedgecentrality}. In detail, we will show that if we perform $\rho =
O(|V| \log |V|$) iterations then the error we make by replacing $L^\kappa(e)$ with
$\hat{L}^\kappa(e)$ is no greater than $\frac{1}{|V|}$.

To prove this result we need the following, preliminary Theorem, known as {\em Hoeffding inequality}:

\begin{theorem}{(Hoeffding inequality)}
{\em Let $X_1, \ldots, X_n$ be independent random variables. Assume that, for each $i$ such that $1
\leq i \leq n$, the random variable $X_i$ ranges in the real interval $\left[ a_i,b_i \right]$. Let
$\overline{X} = (X_1 + \cdots + X_n)/n$. For any $t \geq 0$ we have:
\begin{equation}
	\label{eqn:hoeffdinggeneral}
	\Pr(|\overline{X} - \mathrm{E}[\overline{X}]| \geq t) \leq 2\exp \left( -
	\frac{2t^2n^2}{\sum_{i=1}^n (b_i - a_i)^2} \right)
\end{equation}}
\end{theorem}

{\em Proof}. See \cite{Hoeffding63} \begin{flushright} $\Box$ \end{flushright}.

As a special case, let that all the random variable $X_i$ can only assume the values 0 and 1. In
such a case, Equation \eqref{eqn:hoeffdinggeneral} simplifies to

\begin{equation}
\label{eqn:hoeffdingsimpl}
\Pr(|\overline{X} - \mathrm{E}[\overline{X}]| \geq t) \leq 2\exp \left( -
2t^2n\right)
\end{equation}

We are now able to prove our claims.

\begin{theorem}
\label{th:bounds} {\em Let $G = \langle V, E \rangle$ be a network and $\xi > 0$. Assume to run the
WERW-Kpath algorithm on $G$. The following conditions hold true:

\begin{enumerate}

\item If we set $\rho = \frac{1}{2}\xi^{-2}$, then there exists a constant $\overline{C}$ such
    that $\Pr\left(\left|\hat{L}^\kappa(e) - L^\kappa(e)\right| \geq \xi \right) \leq
    \overline{C}$.

\item There exist two constants $\alpha
> 0$ and $\beta > 0$ such that, if we fix $\rho = \frac{\alpha}{2}|V|^{\beta} \log
|V|$ and $\xi = |V|^{-\frac{\beta}{2}}$, for each edge $e \in E$ we have:
$$ \Pr\left(\left|\hat{L}^\kappa(e) - L^\kappa(e)\right| \geq \xi \right)
    \leq \frac{2}{|V|^{\alpha}}
$$

\end{enumerate}
}
\end{theorem}

{\em Proof}. By Definition \ref{def:kpathedgecentrality}, the edge centrality $L^\kappa(e)$ of an
edge $e$ is defined as follows

\begin{equation}
L^\kappa(e) = \sum_{s\in V}\Pr(e,s)
\end{equation}

being $\Pr(e,s)$ the probability of selecting the edge $e$ starting from the source node $s$.

By the definition of conditional probability we can write

\begin{equation}
\label{eqn:edgecentralityformula}
L^\kappa(e) = \sum_{s\in V}\Pr(e|s)\Pr(s)
\end{equation}

being $\Pr(s)$ the probability that $s$ is the source vertex. Let us now analyze the output
generated by the WERW-Kpath algorithm.

Since the WERW-Kpath algorithm performs $\rho$ iterations, we will first focus on the result
produced in a given iteration, say the $\ell$-th iteration with $1 \leq \ell \leq \rho$.

During the $\ell$-th iteration, a simple random walk of at most $\kappa$ edges is generated. The
edges composing the random walk are selected one-by-one and we will say that we are in the $i$-th
{\em trial} if $i-1$ edges have been already selected.

Let us define the random variable $X_{is}(e)$ as follows

$$
X_{is}(e) = \left\{
			\begin{aligned}
				& 1 \quad  \mbox{if $e$  has been selected at the $i$-th trial {\em and} $s$ is the source vertex}\\
        & 0 \quad \mbox{otherwise.}
			\end{aligned}\right.
			$$

Define now the random variable $Y(e)$ as follows

$$
Y(e) = \sum_{i = 1}^{\kappa}\sum_{s \in V} X_{is}(e)
$$

The variable $Y(e)$ is equal to 1 if $e$ has been selected and 0 otherwise. In fact, independently of
the starting vertex $s$, an edge $e$ can be selected at most one time in all trials (otherwise the
path would pass through it more than once).

By taking the expectation of $Y(e)$ we get

$$
E[Y(e)] = E[\sum_{i = 1}^{\kappa}\sum_{s \in V} X_{is}(e)] = \sum_{i = 1}^{\kappa}\sum_{s \in V} E[X_{is}(e)]
$$

Since $X_{is}(e)$ is an indicator variable, we have that $E[X_{is}(e)] = \Pr(X_{is}(e) = 1)$ (see \cite{Cormen*01} for further details), and therefore

$$
E[Y(e)] = \sum_{i = 1}^{\kappa}\sum_{s \in V} \Pr(X_{is}(e) = 1)
$$

Let us denote as $\Pr(e,i,s) = \Pr(X_{is}(e) = 1)$ and, by Bayes' rule, we have that $\Pr(e,i,s) =
\Pr(e,i|s)\Pr(s)$. We obtain

$$
E[Y(e)] = \sum_{i = 1}^{\kappa}\sum_{s \in V} \Pr(e,i,s)  = \sum_{i = 1}^{\kappa}\sum_{s \in V} \Pr(e,i|s)\Pr(s)
$$

In the WERW-Kpath algorithm, we have that $\Pr(s)$ is equal to $\delta(s)$. By changing the order of the double sum we get

$$
E[Y(e)] = \sum_{s \in V}\left(\sum_{i = 1}^{\kappa} \Pr(e,i|s)\right)\Pr(s)
$$

Let us focus on the term $\sum_{i = 1}^{\kappa} \Pr(e,i|s)$. The WERW-Kpath algorithm generates a
simple random walk, and, therefore, if $e$ is selected in a trial, say $i_1$, it can not be
selected in another trial $i_2$ such that $i_1\neq i_2$. By summing over all indices $i = 1 \ldots
\kappa$, we obtain that $\sum_{i = 1}^{\kappa} \Pr(e,i|s)$ is the probability of selecting $e$
starting from $s$ as the source vertex in an arbitrary trial. As a consequence, $\sum_{i =
1}^{\kappa} \Pr(e,i|s) = \Pr(e|s)$. We can then rewrite $E[Y(e)]$ as

$$
E[Y(e)] = \sum_{s \in V}\Pr(e|s)\Pr(s)
$$

By Equation \eqref{eqn:edgecentralityformula}, $\sum_{s \in V}\Pr(e|s)\Pr(s)$ is equal to
$L^\kappa(e)$, and, therefore $L^\kappa(e) = E[Y(e)]$.

This means that, in a single run of the WERW-Kpath algorithm, the weight associated with $e$ is a
random variable distributed as $Y(e)$ and whose expectation coincide with $L^\kappa(e)$. This
reasoning, of course, holds for any run $\ell$ such that $1 \leq \ell \leq \rho$. Therefroe, the
weight associated with $e$ in the $\ell$-th iteration is a random variable  $Y_{\ell}(e)$.

After completing $\rho$ iterations the algorithm returns, for each edge $e$, the value
$\hat{L}^\kappa(e) = \frac{1}{\rho}\sum_{\ell = 1}^{\rho}Y_{\ell}(e)$. Here $\hat{L}^\kappa(e)$ is
a random variable whose expectation is equal to $L^\kappa(e)$ because $E[\hat{L}^\kappa(e)] =
E[\frac{1}{\rho}\sum_{\ell = 1}^{\rho}Y_{\ell}(e)] = \frac{1}{\rho}\sum_{\ell =
1}^{\rho}E[Y_{\ell}(e)]= \frac{1}{\rho}\sum_{\ell = 1}^{\rho} L^\kappa(e) = \frac{1}{\rho} \rho
L^\kappa(e) = L^\kappa(e)$.

In order to compute how much $\hat{L}^\kappa(e)$ differs from its expectation we can apply the
Hoeffding inequality  as in Equation \eqref{eqn:hoeffdingsimpl}

$$
\Pr\left(\left| \hat{L}^\kappa(e) - L^\kappa(e) \right| \geq \xi\right) \leq 2\exp(-2\rho\xi^{2})
$$

If we set $\rho = \frac{1}{2}\xi^{-2}$, the previous equation simplifies to

$$
\Pr\left(\left| \hat{L}^\kappa(e) - L^\kappa(e) \right| \geq \xi\right) \leq 2\exp(-2\frac{1}{2}\xi^{-2}\xi^{2}) = 2\exp(-1)
$$

By setting $\overline{C} = 2\exp(-1)$ we get the proof for the Part (1) of Theorem \ref{th:bounds}.

As for Part (2), if we fix $\xi = |V|^{-\frac{\beta}{2}}$ and $\rho = \frac{\alpha}{2}|V|^{\beta}
\log |V|$ we get

$$
2\exp(-2\rho\xi^{2}) = 2\exp(-2 \frac{\alpha}{2}|V|^{\beta} \log|V|
|V|^{-\beta}) = 2\exp(-\alpha\log|V|) = 2\frac{1}{|V|^{\alpha}}
$$

and this ends the proof. \begin{flushright} $\Box$\end{flushright}

\bigskip

We can use Theorem \ref{th:bounds} to relate the number of iterations WERW-Kpath has to carry
out with the approximation error it incurs. This is encoded in the following corollary:

\begin{corollary}
\label{cor:bounds} {\em Let $G = \langle V, E \rangle$ be a network. According to the notation
introduced in Theorem \ref{th:bounds}, if we set $\alpha \simeq 1$ and $\beta \simeq 1$, we need to
perform $\rho \simeq |V| \log|V|$ iterations in order to have
$$
\Pr\left(\left| \hat{L}^\kappa(e) - L^\kappa(e) \right| \geq \frac{1}{\sqrt{|V|}}\right) \leq \frac{2}{|V|}
$$
}
\end{corollary}

{\em Proof}. The proof is straightforward by applying Theorem \ref{th:bounds}-Part (2), with
 $\alpha \simeq 1$ and $\beta \simeq 1$. \begin{flushright} $\Box$\end{flushright}

Corollary \ref{cor:bounds} provides us a nice result: in fact, if we perform a number of iterations
in the order of magnitude of $O(|V| \log|V|)$ then the possibility that $\hat{L}^\kappa(e)$ differs
from the actual value $L^\kappa(e)$ more than $\frac{1}{\sqrt{|V|}}$ is less than $\frac{2}{|V|}$.
In real networks $|V|$ is quite large (often in the order of millions). For example, in a network constituted by one million nodes,
the probability that edge centrality values returned by the WERW-Kpath algorithm deviate from the
actual ones more than $10^{-3}$ is less than $10^{-6}$. The consequence is that our algorithm
provides a good trade-off between accuracy and scalability and, therefore, it is fully applicable
in real life scenarios.

To make computation more robust, however, in our experiments we set $\rho = O(|E|)$ and, then, the
worst-case time complexity of the WERW-Kpath algorithm amounts to $O(\kappa|E|)$.

\section{Applying $\kappa$-path edge centrality to find communities}
\label{sec:communitydetection}

In this section we describe how to use the weights produced by the WERW-Kpath algorithm find
communities in networks. We point out that, in principle, our algorithm can be used in conjunction
with any existing \emph{community detection} algorithm. However, due to space limitation, we focus on
three algorithms, namely the Louvain method (LM) \cite{blondel2008fast}, COPRA
\cite{gregory2007algorithm} and OSLOM \cite{lancichinetti2011finding}.

We focused on these algorithms because they show many interesting properties. In detail, the
Louvain method is perhaps one of the best algorithms in terms of accuracy and computational costs.
COPRA is able to find both overlapping and non-overlapping communities and finally, OSLOM is able
to provide a high level of flexibility in the sense that it allows to manage both directed and
undirected graphs, to find overlapping and non-overlapping communities and, finally, to generate a
hierarchy of communities.

In the following we shall describe each of these algorithms in detail.

\subsection{Louvain Method - LM}
\label{sub:LM}

The Louvain method (LM) has been proposed in 2008 by Blondel et al. \cite{blondel2008fast} and it
is perhaps one of the most popular algorithms in the field of community detection. This popularity
derives by the fact that LM provides excellent performance even if the networks to process are very
large. LM consists of two stages which are iteratively repeated. The input of the algorithm is a
weighted network $G = \langle V, E, W \rangle$ being $W$ the weights associated with each
edge\footnote{Of course, in case of unweighted graphs, $W$ is the adjacency matrix of $G$.}. The
modularity is defined as in Equation \eqref{eqn:qmodexp}, in which $A_{ij}$ is the weight of the
edge linking $i$ and $j$ and $k_i$ (resp., $k_j$) is the sum of the edges incident onto $i$ (resp.,
$j$).

Initially, each vertex $i$ will form a community and therefore, there are as many communities as
the vertices in $V$. After that, for each vertex $i$, LM considers the neighbors of $i$; for each
neighboring vertex $j$, LM computes the {\em gain of modularity} that would take place by removing
$i$ from its community and placing it in the community of $j$. The vertex $i$ is placed in the
community for which this gain achieves its maximum value. If it is not possible to
achieve a positive gain, the vertex $i$ will remain in its original community. This process is
applied repeatedly and sequentially for all the vertices until no further improvement can be
achieved. This ends the first phase.

The second step of LM generates a new weighted network $G^{'}$ whose vertices coincide with the
communities identified during the first step. The weight of the edge linking two vertices $i^{'}$
and $j^{'}$ in $G^{'}$ is equal to the sum of the weights of the edges between the vertices in the
communities of $G$ corresponding to $i^{'}$ and $j^{'}$. Once the second step has been performed,
the algorithm re-applies the first step. The two steps are repeated until there are no changes in the obtained
community structure.

LM has three nice properties: {\em (i)} It is a {\em multi-level} algorithm, i.e., it
generates a hierarchy of communities and the $k$-th level of the hierarchy corresponds to the set
of communities found after $k$ iterations of the algorithm. {\em (ii)} The most time expensive
component of the algorithm is the first step and, in particular, the evaluation of the gain the
algorithm could attain by moving a vertex from a community to another one. However, an efficient
formula to quickly compute such a gain has been provided by the authors. {\em (iii)} In the first stage, the algorithm
sequentially scans all the vertices and, for each vertex $i$ it computes the gain achieved by
moving $i$ from its current community to one of the communities of its neighboring vertices.
Therefore, LM is non deterministic because, depending on the ordering of vertices, LM could produce
different results. Experimental trials show that the vertex ordering has no effects on the values
of modularity. However, different vertex orderings could impact on the computational costs of the
algorithm.

\subsection{COPRA}
\label{sub:COPRA}

The COPRA (Community Overlap PRopagation Algorithm) algorithm relies on a {\em label propagation
strategy} proposed for the first time by Raghavan, Albert and Kumara in \cite{raghavan2007near}.
COPRA works in three stages: {\em (i)} Initially, each vertex $v$ is labeled with a set of pairs
$\langle c,b \rangle$, being $c$ a community identifier and $b$ ({\em belonging coefficient}) a
coefficient indicating the strength of the membership of $v$ to the community $c$; belonging
coefficients are also {\em normalized} so that the sum of all the belonging coefficients associated
with $v$ is equal to 1. Initially, the community associated with a vertex coincide with the vertex
itself and the belonging coefficient is 1. {\em (ii)} Then, repeatedly, $v$ updates its label so that
the set of community identifiers associated with $v$ is put equal to the union of the
community identifiers associated with the neighbors of $v$; after that, the belonging coefficients
are updated according to the following formula
$$
b_i(c,v) = \frac{\sum_{w \in N(v)}b_{i-1}(c,v)}{|N(v)|}
$$

being $N(v)$ the set of neighbors of $v$ and $b_{i}(c,v)$ the belonging coefficient associated with
$v$ at the $i$-th iteration. At each iteration, all the pairs in the label of $v$ having a
belonging coefficient less than a threshold are filtered out; in such a case the membership of $v$
to one of the deleted communities is considered not strong enough. It is possible that all the
pairs in a vertex label have a belonging coefficient less than the threshold. In such a case, COPRA
retains only the pair that has the greatest belonging coefficient and deletes all the others.
Finally, if more than one pair has the same maximum belonging coefficient, below the threshold,
COPRA selects at random one of them and this makes the algorithm non-deterministic. After deleting
pairs from the vertex label, the belonging coefficients of each remaining pair are re-normalized so
that they sum to 1. A stopping criterium ensures COPRA ends after a finite number of steps. In such
a case, the set of community identifiers associated with $v$ identify the communities to which $v$
belongs to.

\subsection{OSLOM}
\label{sub:OSLOM}

OSLOM (Order Statistics Local Optimization Method) is a {\em
multi-purpose} technique that aims at managing directed and undirected graphs as well as weighted
and unweighted graphs. OSLOM is also able to detect overlapping communities and to build
hierarchies of clusters.

The strategy to discover clusters in a graph $G$ is as follows: at the beginning a vertex $i$ is
selected at random and it forms the first cluster $C = \{i\}$. After that, the $q$ {\em most
statistically significant} vertices in $G$ are identified and added to $C$. Here $q$ is a random
number and the significance of a vertex $v$ is a parameter indicating the likelihood that $v$ can
be inserted in $C$. To formally define the statistical significance, OSLOM considers a {\em random
null model}, i.e., a class of networks without community structure. A network $G^{'}$ in the random
null model is generated by first copying all the vertices of $G$ in $G^{'}$. After that, multiple
pair of edges in $G^{'}$ are selected at random and an edge is drawn between them. Due to this
procedure, given a vertex $v$ in $G$, there will exist a vertex $v^{'}$ in $G^{'}$ corresponding to
$w$. Analogously, given a subgraph $C$ in $G$, there will be a subgraph $C^{'}$ in $G^{'}$
corresponding to $C$ such that each vertex in $C^{'}$ corresponds to a vertex in $C$. The null model
{\em is expected} not to have a community structure and, therefore, it can be used as a
benchmark to understand if a subgraph $C$ in $G$ is a community and to define the statistical
significance of a vertex $v$ to $C$. In particular, we count the number $l_1$ of vertices linking
$v$ with vertices in $G$; after that, we consider the vertex $v^{'}$ corresponding to $v$ in
$G^{'}$ and we count the number $l_2$ of edges linking $v^{'}$ with vertices residing in $C^{'}$.
If $l_1 > l_2$ we guess that $v$ is significant to $C$ (and can be included in it).

A community $C$ can be associated with a {\em score} representing its {\em quality}; the score of a
cluster $C$ indicates to what extent $C$ contains vertices which have a high statistical
significance with it. The main idea of OSLOM is to {\em progressively} add and remove vertices
within $C$ so that to improve its score; this procedure is called {\em clean-up}.

The whole process introduced above is repeated several times starting from different nodes in
order to explore different regions of $G$. This yields a final set of clusters that may overlap.

\section{Experimental Results}
\label{sec:experiments}

In this section we describe the experiments we carried out to assess the performance of the
WERW-Kpath algorithm and whether its usage is beneficial to raise the quality of a community detection
algorithm.

The WERW-Kpath algorithm has been implemented in Java 1.6 and the prototype is freely available at
the following URL\footnote{ http://www.emilio.ferrara.name/werw-kpath/}.
To perform our tests, we considered 9 datasets whose features are reported in Table
\ref{tab:datasets}.

Dataset 1 is a directed network depicting the voting system of Wikipedia for the elections of January 2008.
Datasets 2--5 represent the undirected networks of Arxiv\footnote{Arxiv (http://arxiv.org/) is an online archive for scientific preprints in the fields of Mathematics, Physics and Computer Science, amongst others.} papers in the field of, respectively, High Energy Physics (Theory), High Energy Physics (Phenomenology), Astro Physics and Condensed Matter Physics, as of April 2003.
Dataset 6 represents a directed network of scientific citations among papers belonging to the Arxiv High Energy Physics (Theory) field.
Dataset 7 represents the directed email communication network of the Enron organization as of 2004, originally made public by the Federal Energy Regulatory Commission during its investigation.
Dataset 8 describes a small sample of the Facebook network, representing its directed friendship graph.
Finally, Dataset 9 depicts a large fragment of the Facebook undirected social graph (mutual friendship relations) as of 2010.

\begin{table*}[!ht]
	\small \centering
	\begin{tabular}{|c l c c c c c|}
		\hline \hline
		N.	&	Network 			&	No. nodes	&	No. edges & Directed 	& Type							&	Ref.\\	
		\hline \hline
		1		&	Wiki-Vote			&	7,115			&	103,689		&	Yes		&	Elections					&	 \cite{leskovec2006sampling}\\
		2		&	CA-HepTh			&	9,877			&	51,971		&	No		&	Co-authors				&	\cite{leskovec2006sampling}\\
		3		&	CA-HepPh			&	12,008		&	237,010		&	No		&	Co-authors				&	\cite{leskovec2006sampling}\\
		4		&	CA-AstroPh	  &	18,772		&	396,160		&	No		&	Co-authors				&	\cite{leskovec2006sampling}\\
		5		&	CA-CondMat		&	23,133		&	186,932		&	No		&	Co-authors				&	\cite{leskovec2006sampling}\\
		6		&	Cit-HepTh			&	27,770		&	352,807		&	Yes		&	Citations					&	\cite{leskovec2006sampling}\\
		7		&	Email-Enron		&	36,692		&	377,662		&	Yes		&	Communications		&	\cite{leskovec2006sampling}\\	
		8		&	Facebook 	    &	63,731		&	1,545,684	&	Yes		&	Online Social	Network &	\cite{viswanath2009evolution}\\
		9		&	SocialGraph 	&	613,497		&	2,045,030	&	No		&	Online Social Network	&	\cite{gjoka2010walking}\\
		\hline \hline
	\end{tabular}
	\caption{Datasets exploited in our tests.}
	\label{tab:datasets}
\end{table*}

Our experiments aim at answering three main research questions:

\begin{itemize}

\item[R1] How much does $\kappa$ impact on the performance of the WERW-Kpath algorithm? From
    Theorem \ref{th:bounds}, we showed that the WERW-Kpath algorithm is convergent, i.e., if
    the number of iterations $\rho$ we carry out grows, then the values returned by the
    algorithm tend to the correct edge centrality values. However, we wonder if wrong choice in
    $\kappa$ may lead to significantly different values of edge centralities. This question
    will be examined in Section \ref{sub:WERWKPathanalysis}.

\item[R2] Is our approach actually capable of improving the modularity of the partitioning
    identified by a community detection algorithm? To answer this question we executed the Louvain
    method, COPRA and OSLOM on the datasets specified above in two configurations: in the
    former we directly applied these algorithms and computed the modularity $Q$ they achieved.
    In the latter, we pre-processed each of these datasets by running our WERW-Kpath algorithm.
    After that, we re-applied LM, COPRA and OSLOM on the modified datasets and re-computed
    the modularity values. The obtained results are discussed in Section
    \ref{sub:assessing-modularity}.

\item[R3] How good are the communities identified by combining the Louvain method, COPRA and
    OSLOM with our algorithms? This task is hard because we should know in advance the actual
    community structure of a network and compare it with that generated by each of these
    algorithms. Unfortunately, such an information is not usually available  for real-life
    networks. Therefore, we used the LFR benchmark \cite{lancichinetti2008benchmark}, a software
    tool proposed to generate artificial networks whose structural features (and in particular
    the communities composing it) can be controlled. We applied the three algorithms with and
    without the pre-processing step by means of WERW-Kpath and, in each configuration, we compared the community
    structure detected by each algorithm with the actual one. To perform such a comparison we
    used a parameter derived from Information Theory known as {\em Normalized Mutual
    Information}. The corresponding results are presented in Section
    \ref{sub:qualityexperiment}.

\end{itemize}

\subsection{Analysis of the WERW-Kpath algorithm}
\label{sub:WERWKPathanalysis}

In this section we study the distribution of edge centrality values computed by the WERW-Kpath
algorithm. In detail, we present the results of two experiments.

\begin{figure}[!ht] \centering
		\includegraphics[width=.45\columnwidth]{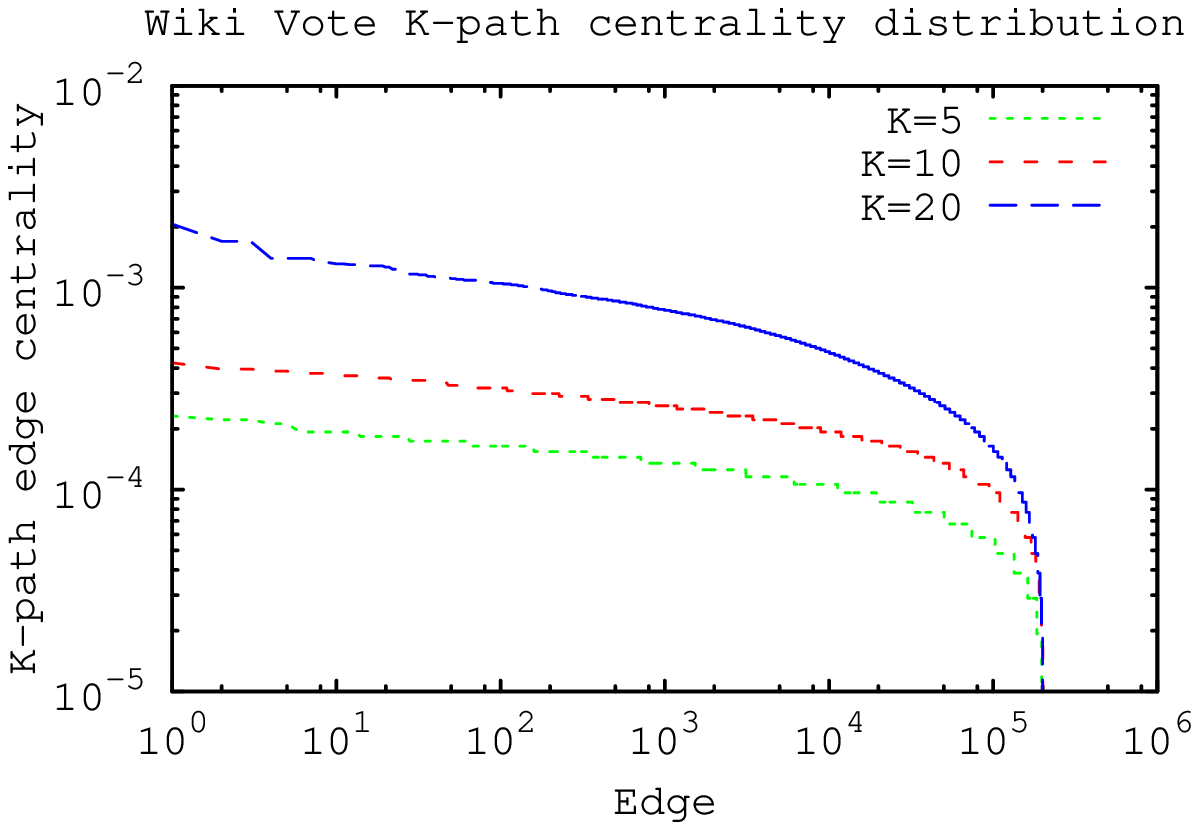}%
		\includegraphics[width=.45\columnwidth]{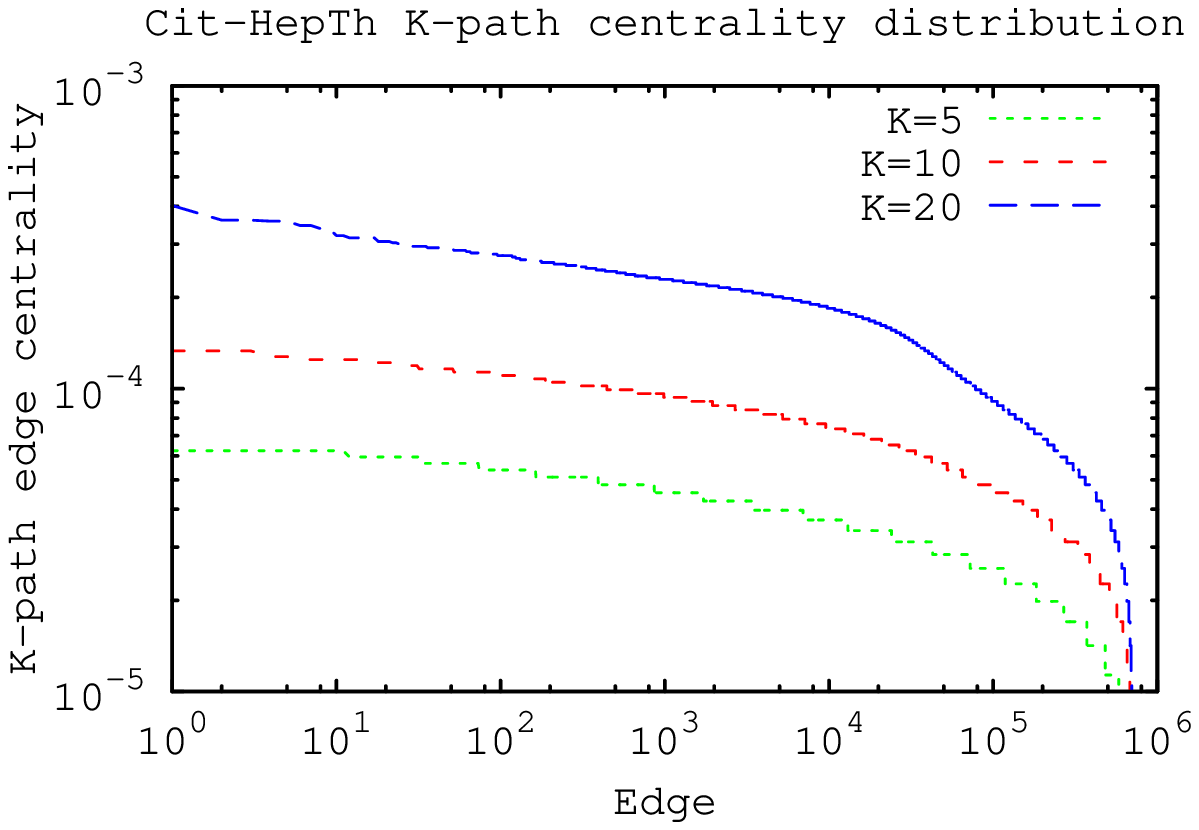}
		\includegraphics[width=.45\columnwidth]{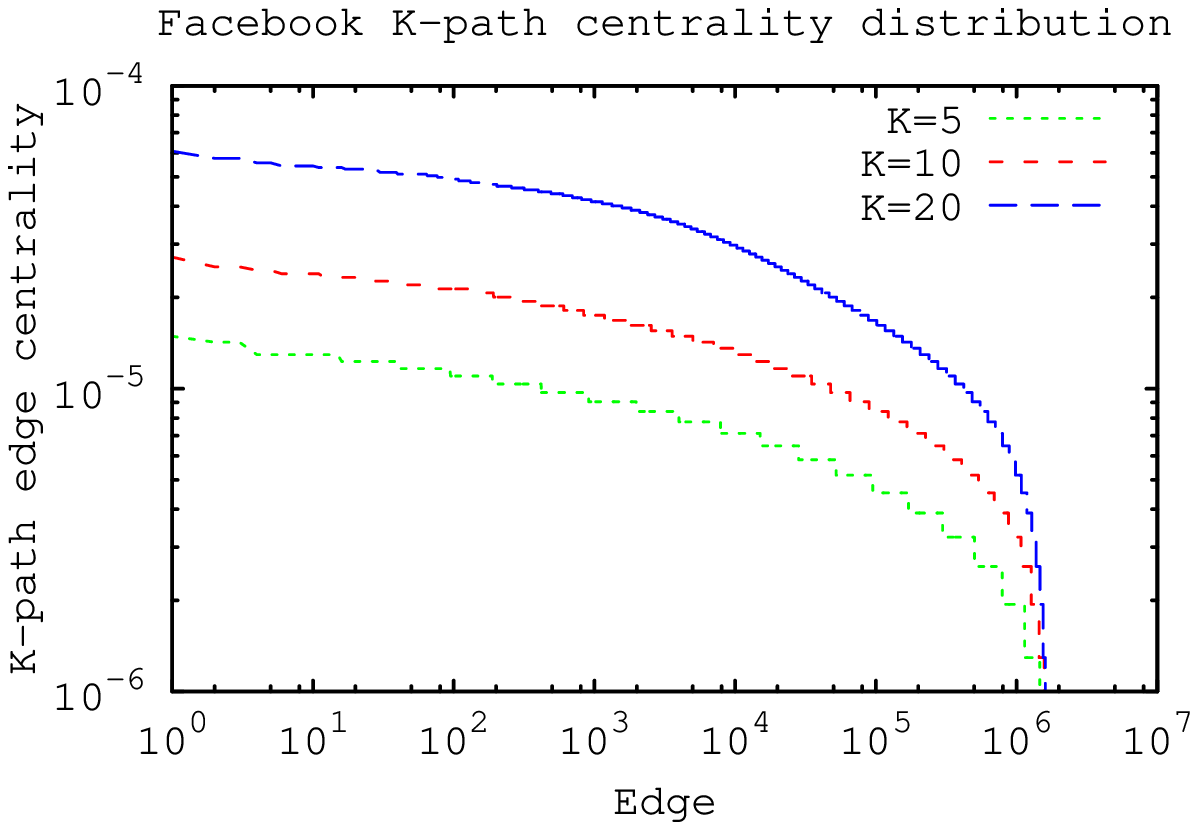}
		\includegraphics[width=.45\columnwidth]{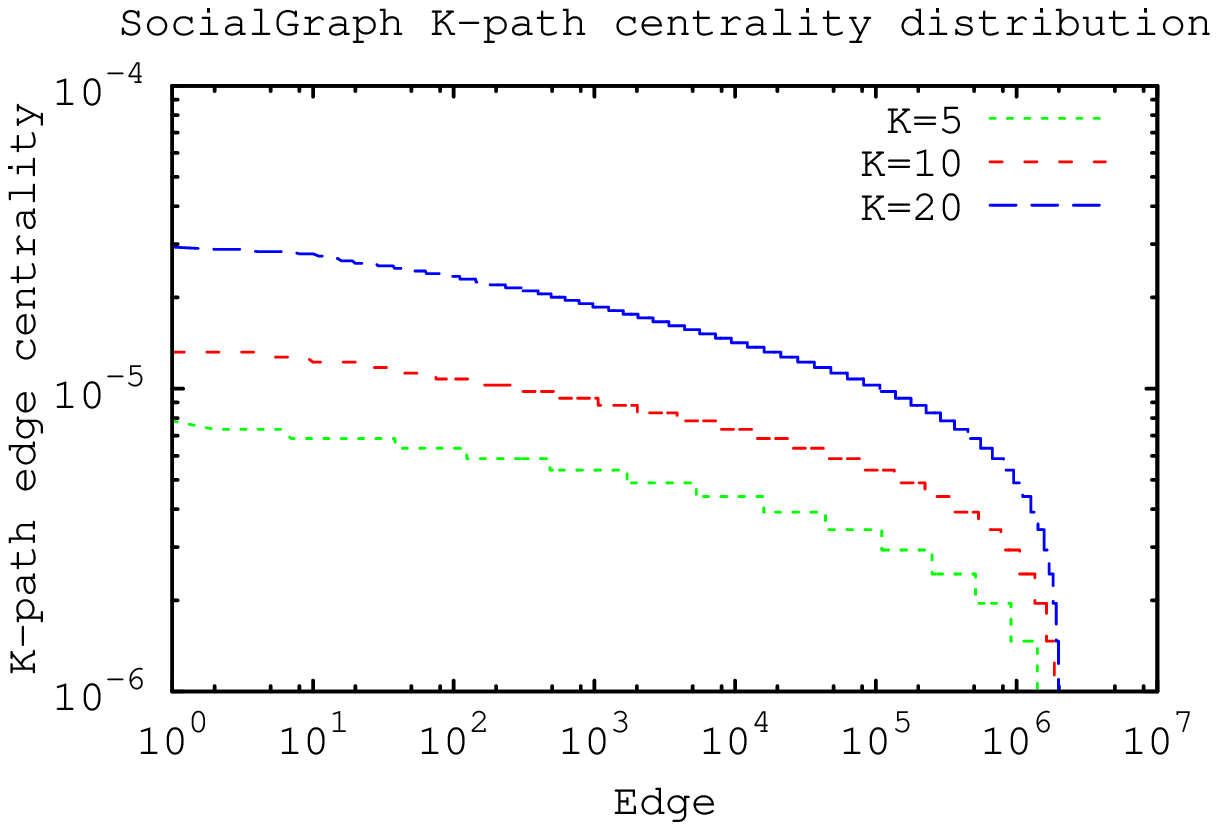}
	\caption{$\kappa$-paths centrality values distribution on different networks.}%
	\label{fig:k-paths}%
\end{figure}

In the first experiment we executed our algorithm four times. In addition, we varied the value of
$\kappa = 5, 10, 20$. We averaged the $\kappa$-path centrality values at each iteration and we
plotted, in Figure \ref{fig:k-paths}, the edge centrality distribution; on the horizontal axis we
reported the identifier of each edge. Due to space limitation, we report only the results we
obtained for four networks of Table \ref{tab:datasets}: a small network (``Wiki-Vote''), two
medium-sized networks (``Cit-HepTh'' and ``Facebook'') and a large-scale network (``SocialGraph'').
Figure \ref{fig:k-paths} exploits a logarithmic scale.

The usage of a logarithmic scale highlights a heavy-tailed distribution for the centrality values.
This means that few edges (which are actually the \emph{most central} edges in a social network)
are frequently selected by the WERW-KPath algorithm and, therefore, their centrality index is
frequently updated. By contrast, many edges are seldom selected and, therefore, their centrality
index is rarely increased.
Heavy-tailed distributions of classic centrality measures, such as the edge betweeness centrality, have been observed in different real-world networks \cite{girvan2002community}.

A further and important result emerging from Figure \ref{fig:k-paths} is that $\kappa$-path edge
centrality, when $\kappa$ is fixed, follows the same trend for all the considered datasets.  This
means that the size of the input dataset does not influence the output of the WERW-Kpath algorithm.

\begin{figure}[!ht] \centering
		\includegraphics[width=.45\columnwidth]{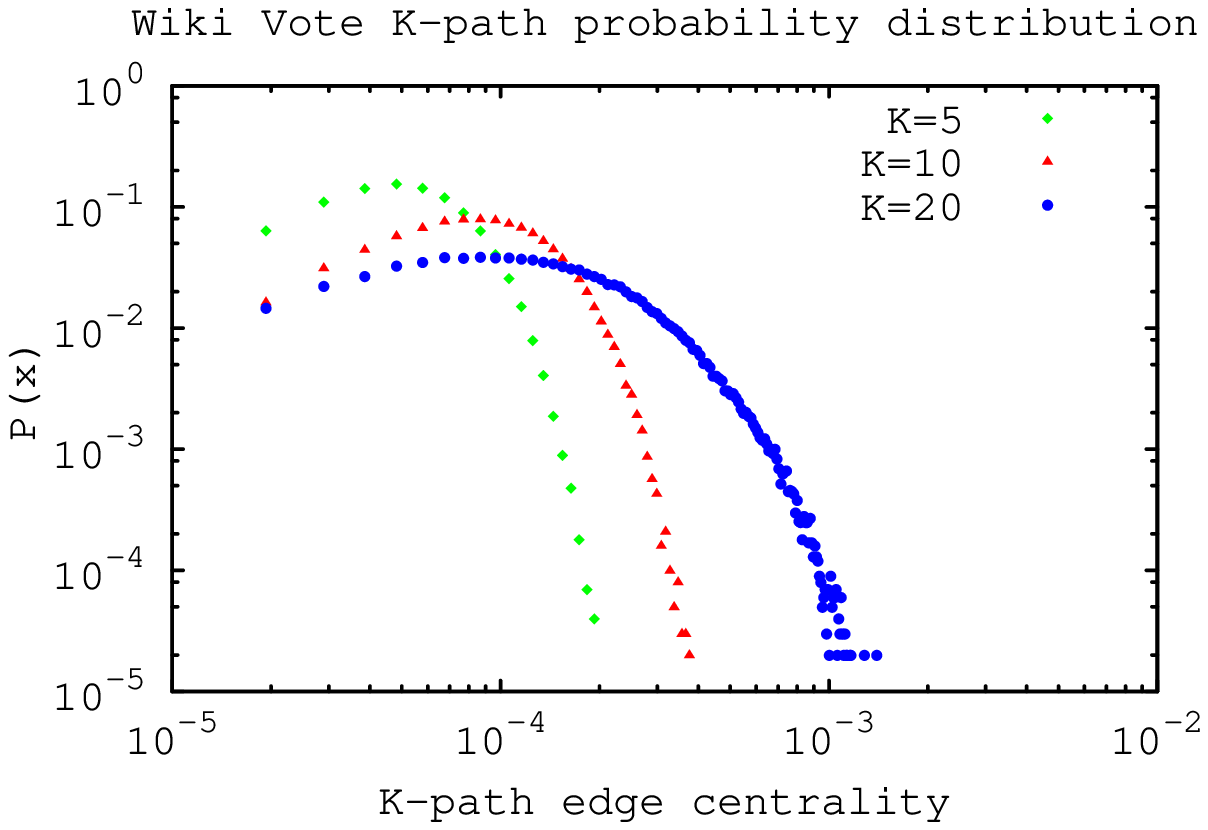}%
		\includegraphics[width=.45\columnwidth]{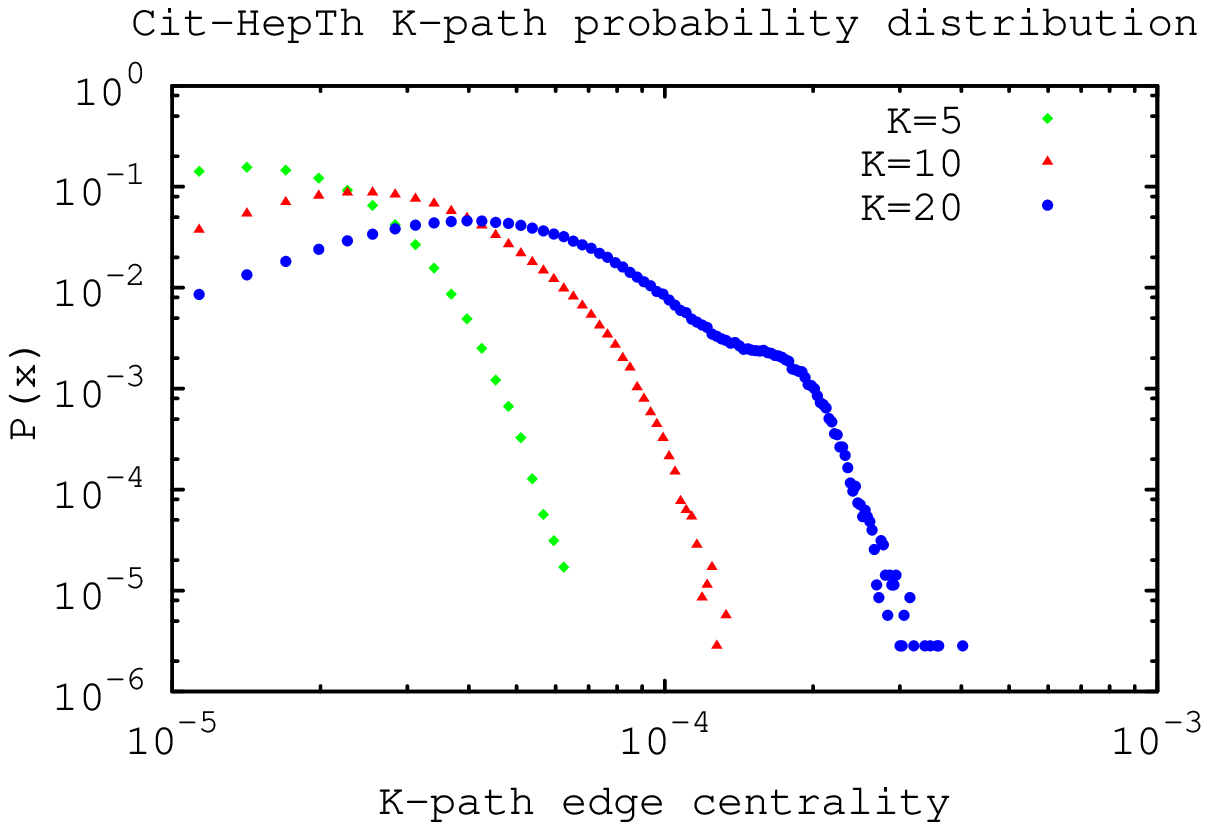}
		\includegraphics[width=.45\columnwidth]{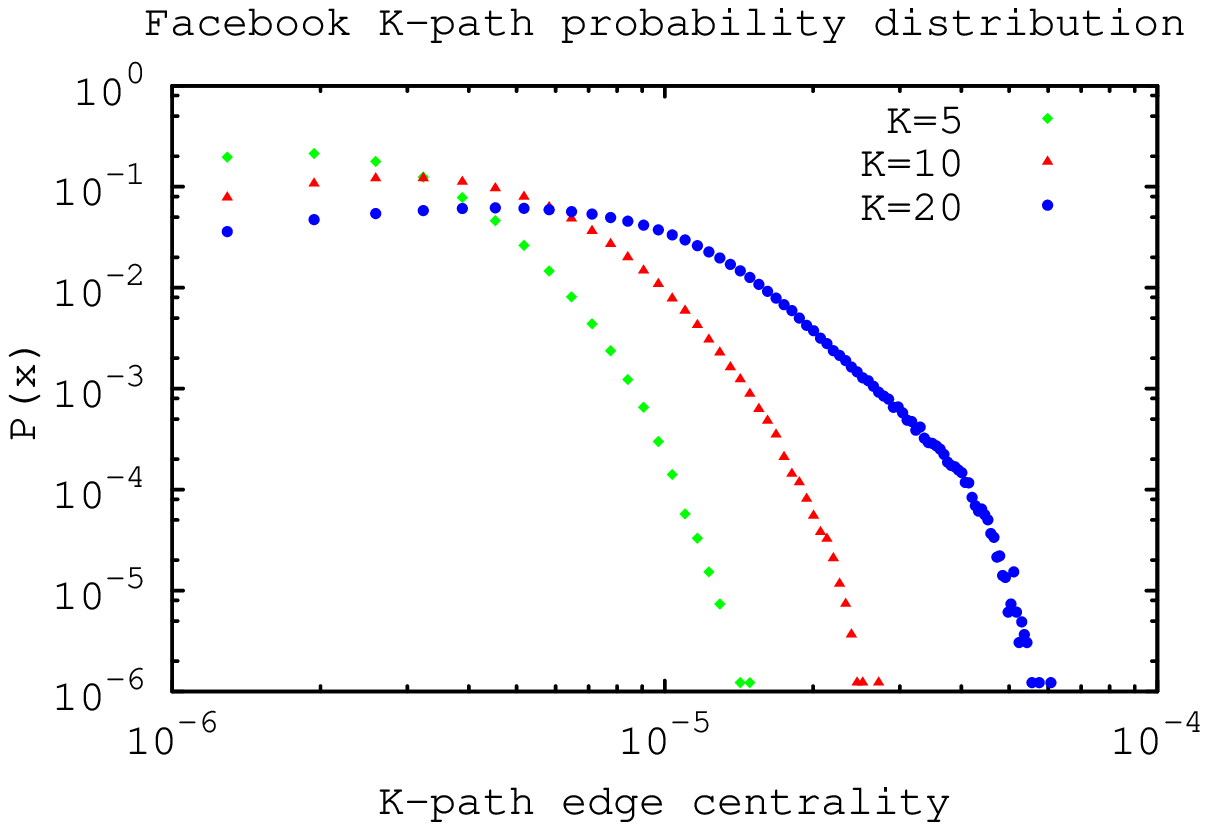}
		\includegraphics[width=.45\columnwidth]{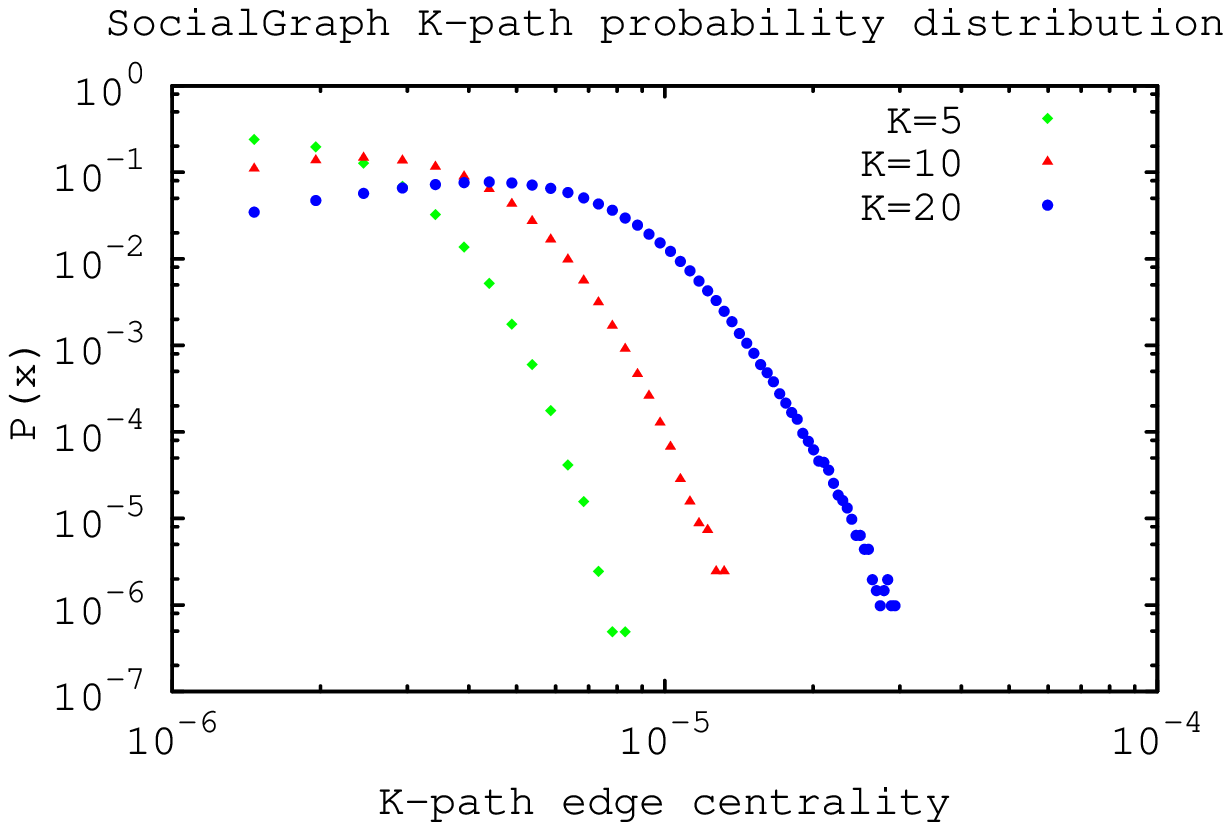}
	\caption{Effect of different $\kappa=5,10,20$ on networks described in Table \ref{tab:datasets}.}%
	\label{fig:k-correlation}%
\end{figure}

In the second experiment, we studied how the value of $\kappa$ impacted on edge centrality. In
detail, we considered the datasets separately and applied the WERW-Kpath algorithm with $\kappa =
5, 10, 20$. After that, for a fixed value of $\kappa$-path edge centrality $\overline{L}$, we
computed the probability $P(\overline{L})$ of finding an edge with such a centrality value.
The corresponding results are plotted in Figure \ref{fig:k-correlation} for the same datasets.
As in the previous case, for each plot we adopted a \emph{log-log} scale.

The analysis of this figure highlights some relevant facts. First of all, the heavy-tailed
distribution in edge centrality emerges in presence of different values of $\kappa$. In other
words, if we use different values of $\kappa$ the centrality indexes may change (see below);
however, as emerges from Figure \ref{fig:k-paths}, for each considered dataset, the curves
representing $\kappa$-path centrality values resemble {\em straight} and {\em parallel} lines with
the exception of the latest part. This implies that, for a fixed value of $\kappa$, say $\kappa =
5$, an edge $\overline{e}$ will have a particular centrality score. If $\kappa$ grows from 5 to 10
and, then, from 10 to 20, the centrality of $\overline{e}$ will be increased by a constant factor.

This leads us to hypothesize that a form of correlation should exist between the values of $L^{\kappa}(e)$ for different values of $\kappa$.
To check whether this hypothesis were true, we performed a further experiment. In detail, we considered the above mentioned datasets and applied {\em twice} our algorithm on each of these datasets with two different values of $\kappa$, say $\kappa_{X}$ and $\kappa_{Y}$. Let us denote as $D_X$ (resp., $D_Y$) the distribution of edge centralities computed when $\kappa = \kappa_X$ (resp., $\kappa = \kappa_Y$). We compared $D_X$ and $D_Y$ and, to this purpose, we computed the Pearson Correlation coefficient $r_{\kappa_X,\kappa_Y}$ of $D_X$ and $D_Y$. The Pearson Correlation Coefficient assumes that a linear relationship exists between $D_X$ and $D_Y$; such an assumption may be false because, in our scenario, we do not know if a linear relationship between $D_X$ and $D_Y$ exists and this could make the process of comparing $D_X$ and $D_Y$ unreliable. For instance, if $D_X$ and $D_Y$ were strongly correlated but both $D_X$ and $D_Y$ would contain some outliers, this would lead to significantly low values of $r_{\kappa_X,\kappa_Y}$ and, in such a case, we should erroneously conclude that $D_X$ and $D_Y$ are weakly correlated. Due to these reasons, to make our analysis more robust, we computed two further metrics: the {\em Spearman's rank correlation coefficient} $\rho_{\kappa_X,\kappa_Y}$ and {\em Kendall's tau rank correlation coefficient} $\tau_{\kappa_X,\kappa_Y}$. Both these parameters are useful to determine whether two variables (in our case $D_X$ and $D_Y$) are related by a {\em monotonic function}, i.e., they are useful to identify to what extent when the former variable tend to increase the latter tends to increase too or to decrease.

The Spearman's rank correlation coefficient is computed by converting the edge centralities into {\em rank values} (in such a way as to the edge with the highest centrality is ranked as first); subsequently, the Pearson Correlation Coefficient is computed on the rank values. The Spearman's rank correlation coefficient ranges in $[-1,1]$.

The definition of the Kendall's tau rank correlation coefficient is slightly complex; to this purpose, let us consider a pair of edges $e_i$ and $e_j$ and let us denote as $L^{\kappa_X}(e_i)$ (resp., $L^{\kappa_X}(e_j)$) and $L^{\kappa_Y}(e_i)$ (resp., $L^{\kappa_Y}(e_j)$) their centralities computed when $\kappa = \kappa_X$ and  $\kappa = \kappa_Y$. The pairs $\langle L^{\kappa_X}(e_i), L^{\kappa_X}(e_j) \rangle$ and $\langle L^{\kappa_Y}(e_i), L^{\kappa_Y}(e_j) \rangle$ are said {\em concordant} if both $L^{\kappa_X}(e_i) > L^{\kappa_Y}(e_i)$ and $L^{\kappa_X}(e_j) > L^{\kappa_Y}(e_j)$ or $L^{\kappa_X}(e_i) < L^{\kappa_Y}(e_i)$ and $L^{\kappa_X}(e_j) < L^{\kappa_Y}(e_j)$. The same pairs are said {\em discordant} if both $L^{\kappa_X}(e_i) > L^{\kappa_Y}(e_i)$ and $L^{\kappa_X}(e_j) < L^{\kappa_Y}(e_j)$ or $L^{\kappa_X}(e_i) < L^{\kappa_Y}(e_i)$ and $L^{\kappa_X}(e_j) > L^{\kappa_Y}(e_j)$. The Kendall's tau coefficient is equal to the ratio of the difference between the number of concordant and discordant pairs to the total number of pairs in distributions $D_X$ and $D_Y$ and it ranges in the real interval $[-1,1]$.

In Table \ref{tab:correlation} we report the outcomes of our experiment. Due to space limitations, we report four out of the nine datasets reported in Table \ref{tab:datasets} -- the same of the previous experiment, namely ``Wiki-vote'', ``Cit-HepTh'', ``Facebook'' and ``SocialGraph''. From Table \ref{tab:correlation}, it emerges a strong agreement among the values returned by $r_{\kappa_X,\kappa_Y}$,	$\rho_{\kappa_X,\kappa_Y}$ and $\tau_{\kappa_X,\kappa_Y}$. In detail, all the metrics introduced above clearly indicate that there is a {\em strong} and {\em positive correlation} between $D_X$ and $D_Y$ for {\em all datasets} and for any pair of values $\kappa_X$ and $\kappa_Y$. Such a result highlights a nice property of our algorithm: the agreement between the two rankings produced for different values $\kappa_X$ and $\kappa_Y$ is the same; as a consequence, the edge having the highest centrality value when $\kappa = \kappa_X$ will be also the edge with the highest centrality value when $\kappa = \kappa_Y$. Therefore, our algorithm is robust against variations in the value of $\kappa$ and we can safely conclude that different values of $\kappa$ do not alter the distribution of edge centralities.

%

As previously observed, if $\kappa$ increases, the centrality index of an edge increases too (or,
at least, it does not decrease). This has an intuitive explanation: if $\kappa$ increases, the
WERW-KPath algorithm manages longer paths and, therefore, the chance that an edge is selected
multiple times increases too. Each time an edge is selected, WERW-Kpath increases its weight by
1 and this increases the edge centrality values. Figure \ref{fig:k-paths} show that if we augment
$\kappa$, the distance between the highest and the lowest centrality value increase too. Therefore,
in presence of low values of $\kappa$, edge centrality indexes tend to edge flatten in a small
interval and it is harder to distinguish high centrality edges from low centrality ones. Vice
versa, in presence of high values of $\kappa$, we are able to better discriminate edges with high
centrality from edges with low centrality.

As a consequence, on one hand, it would be fine to fix $\kappa$ as high as possible. On the other
hand, since the complexity of our algorithm is $O(\kappa |E|)$, large values of $\kappa$
negatively impact on the performance of our algorithm.

A good trade-off (suggested by the experiments showed in this section) is to fix $\kappa = 20$.

\begin{table}
	\small \centering
	\begin{tabular}{|cccccc|}
	\hline \hline
	Network	&	$\kappa_X$	&	$\kappa_Y$	&	$r_{\kappa_X,\kappa_Y}$	&	$\rho_{\kappa_X,\kappa_Y}$	&	$\tau_{\kappa_X,\kappa_Y}$\\
	\hline \hline
	Wiki-vote	&	5	&	10	&	0.9916	&	0.9897	&	0.9569\\
			&	10	&	20	&	0.9765	&	0.9976	&	0.9792\\
			&	20	&	5	&	0.9701	&	0.9925	&	0.9531\\
	Cit-HepTh	&	5	&	10	&	0.9896	&	0.9907	&	0.9611\\
			&	10	&	20	&	0.9861	&	0.9971	&	0.9781\\
			&	20	&	5	&	0.9664	&	0.9924	&	0.9924\\
	Facebook	&	5	&	10	&	0.9874	&	0.9810	&	0.9436\\
			&	10	&	20	&	0.9931	&	0.9952	&	0.9714\\
			&	20	&	5	&	0.9848	&	0.9858	&	0.9379\\
	SocialGraph	&	5	&	10	&	0.9803	&	0.9772	&	0.9366\\
			&	10	&	20	&	0.9924	&	0.9910	&	0.9608\\
			&	20	&	5	&	0.9834	&	0.9811	&	0.9288\\
	\hline \hline	
	\end{tabular}
	\label{tab:correlation}
	\caption{Pearson ($r_{\kappa_X,\kappa_Y}$), Spearman ($\rho_{\kappa_X,\kappa_Y}$) and Kendall tau ($\tau_{\kappa_X,\kappa_Y}$) rank correlation coefficients for pairs of distributions in Figure \ref{fig:k-correlation} (the \emph{p-value} for all values is less than $10^{-10}$).}
\end{table}

\subsection{Assessing the Modularity}
\label{sub:assessing-modularity}

In this section we analyze the modularity of the partitions achieved by Louvain method, COPRA and
OSLOM with and without the support of the WERW-Kpath algorithm.

The experiment has been carried out as follows: in a first stage we considered the original datasets
(which can be regarded as unweighted graphs) and applied the Louvain method, COPRA and OSLOM on
them. We computed the modularity achieved by each algorithm on each of these
datasets. To make the notation simple we shall use the labels LM UW, CP UW and OS UW to denote,
respectively, the Louvain method, COPRA and OSLOM applied on the unweighted, original, datasets.

In the second stage we pre-processed each of the 9 datasets reported in Table \ref{tab:datasets} by
applying the WERW-Kpath algorithm with $\kappa = 20$. Therefore, each graph was transformed into a
weighted graph in which the weight of an edge was equal to its edge centrality. We applied again
the three algorithms\footnote{Clearly, in this case we adopted those versions of the algorithms
designed for weighted networks.} on the aforementioned datasets after adopting our strategy and we
computed the achieved modularity. Similarly to previous case, we shall use the label LM W, CP W and OS
W to denote, respectively, the Louvain method, COPRA and OSLOM applied on weighted, pre-processed,
datasets.

The corresponding results are reported in Table \ref{tab:networkmodularity}. From the analysis of
this table we conclude that:

\begin{enumerate}

\item As for LM, the usage of the WERW-Kpath algorithm always yields better results than
    using the original LM alone. In particular, the improvement of modularity is up 17.3\%. It
    is interesting to observe that in the large-scale dataset (i.e., ``SocialGraph'') a very
    high value of modularity was achieved by applying LM alone ($Q = 0.891$). Neverthless, the
    WERW-KPath algorithm gives room for a further improvement ($Q = 0.912$). This result is
    interesting because WERW-Kpath can yield relevant improvements also on large datasets
    for which the optimization of modularity becomes increasingly hard.

\item As for COPRA, there are two datasets (namely ``CA-HepTh'' and ``CA-CondMat'') in which the
    performance of the community detection algorithm in conjunction with the WERW-Kpath algorithm produce a worse
    modularity than that achieved by COPRA alone. In all other cases, we report an increase of
    $Q$ ranging from 1.37\% (``Facebook'' dataset) to 9.02\% (``Cit-HepTh'' dataset). The combination
    of COPRA with the WERW-Kpath algorithm seems favorable for networks of medium size but it
    deteriorates for small and large networks. For instance, the increase of $Q$ is around
    1.73\% for the dataset called ``Wiki-Vote'' (7,115 vertices and 103,689 edges) and, as
    previously pointed out, 1.37\% for the ``Facebook'' dataset (63,731 vertices and 1,545,684
    edges). Such an improvement is better for ``SocialGraph'' because we pass from $Q = 0.197$
    achieved by COPRA alone to $Q = 0.203$ with a gain of 3.04\%. In such a case, however, the
    modularity achieved by COPRA is quite low in comparison with that of Louvain method and
    OSLOM and, therefore, such an improvement is not particularly relevant.

\item As for OSLOM, the improvements associated with the adoption of our method is less
    evident. In fact, we can observe that for 6 datasets out of 9 the joint usage of the
    WERW-KPath algorithm with OSLOM produce better results than those we would achieve if we
    would apply OSLOM alone. The improvement of $Q$ ranges from 2.8\% to 7\% (excluding the exceptional +17.4\% for the dataset ``Wiki-Vote''). To explain these
    results we can observe that OSLOM does not target at maximizing the network modularity but
    it relies on the idea that vertices can be ranked according to their likelihood of
    belonging to a community. Therefore, it is not surprising that the network modularity
    achieved by OSLOM is significantly less than that achieved by the Louvain method. However,
    it is worth observing that the gain in modularity deriving from the usage of the WERW-Kpath
    algorithm is {\em almost uniform}: for instance, the increase of $Q$ due to the coupling of
    WERW-Kpath with OSLOM is 7\% for the dataset called ``CA-AstroPh'', 6.21\% for the dataset
    called ``CA-CondMat'', 6.66\% for the dataset called ``Cit-HepTh'' and 4.82\% for the
    dataset called ``SocialGraph''. This implies that the joint usage of WERW-KPath is not
    influenced (or at least is weakly influenced) by the size of the input dataset.

\end{enumerate}

\begin{table*}[!ht]
	\small \centering
	\begin{tabular}{|l c c c c c c|}
		\hline \hline
		Network 				& LM  UW  & LM  W  & CP UW & CP W  & OS UW  & OS W \\	
		\hline \hline
		Wiki-Vote				& 	0.423	& 	\textbf{0.445}	[+5.2\%]	& 0.693 	&	\textbf{0.705}	[+1.7\%]	&	0.316	&	\textbf{0.371}	 [+17.4\%]\\
		CA-HepTh			& 	0.772	& 	\textbf{0.806}	[+4.4\%]	& 0.768	&	\emph{0.649}	[-18.4\%]	&	0.653	&	\emph{0.632}	 [-3.3\%]\\
		CA-HepPh			&	0.656	& 	\textbf{0.760}	[+15.8\%]	& 0.754	&	\textbf{0.777}	[+3.1\%]	&	0.675	&	\emph{0.669}	 [-0.01\%]\\
		CA-AstroPh			&	0.627 	& 	\textbf{0.663}	[+5.7\%]	& 0.577 	&	\textbf{0.614}	[+6.4\%]	&	0.596	&	\textbf{0.638}	 [+7.0\%]\\
		CA-CondMat			& 	0.731	& 	\textbf{0.768}	[+5.1\%]	& 0.616	&	\emph{0.515}	[-19.6\%]	&	0.692	&	\textbf{0.735}	 [+6.2\%]\\
		Cit-HepTh			&	0.642	&	\textbf{0.644}	[+0.1\%]	& 0.665	&	\textbf{0.725}	[+9.0\%]	& 	0.433	&	\textbf{0.462}	 [+6.7\%]\\
		Email-Enron			&	0.602	&	\textbf{0.706}	[+17.3\%]	& 0.768	&	\textbf{0.799}	[+4.0\%]	& 	0.449	& 	\emph{0.432}	 [-4.0\%]\\	
		Facebook 			&	0.626 	& 	\textbf{0.664}	[+6.1\%]	& 0.799	&	\textbf{0.810}	[+1.4\%]	&	0.391	&	\textbf{0.402}	 [+2.8\%]\\
		SocialGraph 			&	0.891	&	\textbf{0.912}	[+2.4\%]	& 0.197	&	\textbf{0.203} 	[+3.0\%]	&	0.456	&	\textbf{0.478}	 [+4.8\%]\\
		\hline \hline
	\end{tabular}
	\caption{Network Modularity of Louvain method, COPRA  and OSLOM with and without our approach ($\kappa = 20$). Improved values are highlighted in bold, reduced values are emphasized and the improvement/loss is reported in brackets.}
	\label{tab:networkmodularity}
\end{table*}

The results reported in this section indicates that some community detection methods (like LM) significantly benefit from our edge weighting strategies; for COPRA and OSLOM our method brings some advantage but the improvement in modularity is less evident than in LM. Such a behavior can be explained in terms of the properties and behavior of the LM, COPRA and OSLOM algorithms.

As emerges from Section \ref{sub:LM}, LM tries to optimize in a greedy fashion the modularity function $Q$ defined in Equation \ref{eqn:qmodexp}. The $Q$ function is the sum of terms of the form $\Delta_{ij} = \left(A_{ij} - \frac{k_i \cdot k_j}{2m}\right) \delta(C_i,C_j)$. Let us consider a pair of vertices $i$ and $j$ and assume that an edge linking them exists. By exploiting LM in conjunction with our WERW-Kpath algorithm, $A_{ij}$ is a real number in [0,1]; if the WERW-Kpath algorithm is not applied, $A_{ij} \in{0,1}$ and it equals 1 if and only if there is an edge linking $i$ and $j$. In such a case the term $\Delta_{ij}$ reads $\hat{\Delta_{ij}} = \left(1 - \frac{k_i \cdot k_j}{2m}\right) \delta(C_i,C_j)$. Therefore, the less $k_i$ (resp., $k_j$) the higher $\hat{\Delta_{ij}}$: this implies that LM tends to put in the same community vertices at low degree even if such a choice could be not optimal at the global level. By contrast, the usage of non-binary weights on edges provides a higher level of flexibility to the algorithm and this ultimately explains the improvements in the values of $Q$.

As for COPRA and OSLOM, the discussion proposed in Sections \ref{sub:COPRA} and \ref{sub:OSLOM} explains that these algorithms deeply differ each other. Despite these deep differences, they share a relevant similarity: in both of them, a vertex $v$ is assigned to a community depending on the fact that a large part of the neighboring vertices of $v$ belong or not to that community. So, for instance, in COPRA we compute the belonging coefficient of $v$ to a community $C$ depending on the belonging coefficient to $C$ of its neighboring vertices. In OSLOM, we compute the number of links joining $v$ with vertices located inside $C$ and if such a number is higher than that we would expect, we decide to put $v$ in $C$. In both COPRA and OSLOM, the criterium to decide if a vertex has to be included in a community depends on the number of its neighboring vertices belonging to that community and, ultimately, on the number of edges linking $v$ with vertices located in $C$. Therefore, weights on edges have a small (or negligible) influence in deciding to assign a vertex to a community.

\subsection{Quality Assessment}
\label{sub:qualityexperiment}

In this section we analyze the quality of the communities detected by our approach in conjunction
with LM, COPRA and OSLOM.

To assess the quality of the results, we adopted a measure called {\em Normalized Mutual
Information - NMI} proposed by Danon et al. in 2005 \cite{danon2005comparing} which is rooted in
Information Theory. Such a measure assumes that, given a graph $G$, a {\em ground truth} is
available to verify what are the communities (said {\em real communities}) in $G$ and what are
their features. Let us denote as $A$ the true community structure of $G$ and suppose that $G$
consist of $c_A$ communities. Let us consider a community detection algorithm $\cal A$. Let us run
$\cal A$ on $G$ and assume that it identifies a community structure $B$ consisting of $c_B$
communities. We define a $c_A \times c_B$ matrix -- said {\em confusion matrix} -- $CM$ such that
each row of $CM$ corresponds to a community in $A$ whereas each column of $CM$ is associated with a
community in $B$. The generic element $CM_{ij}$ is equal to the number of elements of the real
$i$-th community which are also present in the $j$-th community found by the algorithm. Starting by
this definition, the normalized mutual information is defined as

\begin{equation}
\label{eqn:mutualinformationdef}
NMI(A,B) = \frac{-2\sum_{i = 1}^{c_A}\sum_{j = 1}^{c_B} N_{ij} \log \left(\frac{N_{ij}N}{N_{i\cdot}N_{\cdot j}} \right)}{\sum_{i=1}^{c_A}N_{i \cdot} \log \left(\frac{N_{i \cdot}}{N}\right) + \sum_{j=1}^{c_B}N_{\cdot j} \log \left(\frac{N_{\cdot j}}{N}\right)}
\end{equation}

being $N_{i \cdot}$ (resp., $N_{\cdot j}$) the sum of the elements in the $i$-th row (resp., $j$-th
column) of the confusion matrix. If the algorithm $\cal A$ would work perfectly, then for each
found community $j$, it would exist a real community $i$ exactly coinciding with $j$. In such a
case, it is possible to show that $NMI(A,B)$ is exactly equal to 1 \cite{danon2005comparing}. By
contrast, if the communities detected by $\cal A$ are totally independent of the real communities
(e.g. if we assume to put all the nodes of the network into a single community) then it is possible
to show that the NMI is equal to 0. The NMI, therefore, ranges from 0 to 1 and the higher the
value, the better the algorithm works.

The computation of NMI is however challenging for real-life networks because no {\em ground
truth} is usually available to assess what are the communities in $G$ and what are their features.
Therefore, to perform our tests, we need to consider a set of artificially generated networks whose
structural properties are compliant with those existing in real networks.

A tool for generating artificial networks resembling real ones has been proposed in
\cite{lancichinetti2008benchmark} and it has been exploited in our tests. The user is required to
provide the following parameters to generate artificial networks: {\em (i) Number of Vertices and
Average Vertex Degree}.  The user is allowed to specify the number $N$ of vertices in the network
as well as the average degree $\langle k \rangle$ of each vertex. {\em (ii) Power Law exponent in
vertex degree distribution}. The user specifies a parameter $\gamma$ such that the vertex degree
distribution follows a power law such as $P(k) \propto k^{-\gamma}$. In addition, the average
degree of a vertex is fixed to be equal to $\langle k \rangle$. {\em (iii) Power Law exponent in
community size distribution}. The user specifies a parameter $\beta$  and communities are generated
so that the size of each community (i.e., the number of vertices composing it) follows a power
law defined as $f(x) \propto x^{-\beta}$. The sum of the sizes of all the communities is
constrained to be equal to $N$. In addition, the procedure for generating communities ensures that
any node is included in at least a community, independently of its degree. {\em (iv) Mixing
parameter}. The user specifies a parameter $\mu \in (0,1)$ such that each vertex shares a fraction
$1 - \mu$ of its edges with vertices outside its community and $\mu$ edges with vertices residing
in its community. The parameter $\mu$ is called {\em mixing parameter}. Note that the mixing
parameter assignment $\mu = 0.5$ represents the tipping point beyond which the communities are no
longer defined in the strong sense, that is that each vertex has more neighbors in the community to
which it is assigned, rather than outside.

In our tests we adopted the same configuration reported in \cite{lancichinetti2008benchmark},
i.e.,: {\em (i)} $N=1000$ vertices; {\em (ii)} four pair of values for $\gamma$ and $\beta$ were
considered, namely: $(\gamma,\beta)$ = (2,1), (2,2), (3,1), (3,2); {\em (iii)} three values of
average degree were considered, namely $\langle k \rangle = 15,20,25$; {\em (iv)} six values of
$\mu$ were considered, namely $\mu = 0.1, \dots, 0.6$. This allowed us to generate an overall
number of $4 \cdot 3 \cdot 6 = 72$ artificial test networks.

We computed the NMI achieved by applying \emph{LM}, \emph{COPRA} and \emph{OSLOM}, on the unweighted networks as they were generated by the benchmark, averaging obtained results over 10 runs of each algorithm.
Therefore, we applied the same algorithms to the weighted networks after the adoption of our network weighting strategy, once again averaging results over 10 runs. As suggested in Section \ref{sub:assessing-modularity}, we fixed $\kappa = 20$ to compute edge centralities.

The achieved results are reported in Table \ref{tab:artificial} for $\langle k \rangle = 20$. We also considered other values for average degree, namely $\langle k \rangle = 15$ and $\langle k \rangle = 25$ and the results we obtained were quite similar each other and inline with those obtained when $\langle k \rangle = 20$; therefore, due to space limitation, we report in Table \ref{tab:artificial} only the NMI values obtained when $\langle k \rangle = 20$.

\begin{table}[!ht]
	\small \centering
	\begin{tabular}{|l c c c c c c|}
		\hline \hline
		Method &	$\mu = 0.1$	&	$0.2$	&	$0.3$	&	$0.4$	&	$0.5$	& $0.6$  \\

		\hline \hline
		\multicolumn{7}{|c|}{$\gamma = 2, \beta = 1$}\\
		\hline
		 LM UW 		& 0.917	& 0.853	& 0.769	& 0.732	& 0.591	& 0.486\\
		 LM W 	 	& \textbf{0.931}	& \textbf{0.882}	& \textbf{0.817}	& \textbf{0.789}	& 0.599	& 0.444\\
		 \hline
		 CP UW 	 	& 0.868	& 0.901	& 0.841	& 0.868	& 0.691	& 0.021\\
		 CP W 	 		& 0.879	& 0.900	& \textbf{0.905}	& 0.856	& \textbf{0.846}	& 0.021\\
		 \hline
		 OS  UW	 	& 0.699	& 0.703	& 0.705	& 0.694	& 0.658	& 0.438\\
		 OS  W  	& 0.700	& 0.704	& 0.707	& 0.695	& 0.668	& \textbf{0.501}\\		
		 \hline \hline
		\multicolumn{7}{|c|}{$\gamma = 2, \beta = 2$}\\
		\hline
		 LM UW 		& 0.815	& 0.633	& 0.664	& 0.428	& 0.503	& 0.334\\
		 LM W 	 	& \textbf{0.886}	& \textbf{0.704}	& 0.632	& \textbf{0.519}	& 0.444	& \textbf{0.377}\\
		 \hline
		 CP UW 	 	& 0.849	& 0.893	& 0.838	& 0.809	& 0.726	& 0.028\\
		 CP W 		& \textbf{0.904}	& 0.852	& \textbf{0.886}	& 0.759	& 0.738	& 0.028\\
		 \hline
		 OS UW 	 	& 0.669	& 0.637	& 0.629	& 0.633	& 0.583	& 0.462\\
		 OS W 		& 0.669	& 0.637	& 0.637	& 0.629	& \textbf{0.602}	& \textbf{0.505}\\		
		 \hline \hline
		\multicolumn{7}{|c|}{$\gamma = 3, \beta = 1$ }\\
		\hline
		 LM UW 		& 0.973	& 0.867	& 0.800	& 0.773	& 0.677	& 0.527\\
		 LM W 	 	& 0.978	& 0.872	& 0.806	& 0.739 & \textbf{0.712}	& 0.404\\
		 \hline
		 CP UW 	 	& 0.917	& 0.888	& 0.876	& 0.875	& 0.727	& 0.018\\
		 CP W 	    & 0.927	& 0.892	& \textbf{0.892}	& \textbf{0.913}	& \textbf{0.786}	& 0.018\\
		 \hline
	   OS UW 	 	& 0.741	& 0.711	& 0.700	& 0.712	& 0.708	& 0.299\\
		 OS W  		& 0.741	& 0.712	& 0.700	& 0.712	& 0.711	& \textbf{0.415}\\		
		 \hline \hline
		\multicolumn{7}{|c|}{$\gamma = 3, \beta = 2$}\\
		\hline
		 LM UW 		& 0.936	& 0.788	& 0.692	& 0.563	& 0.532	& 0.411\\
		 LM W 	 	& 0.947	& 0.745	& \textbf{0.749}	& \textbf{0.633}	& \textbf{0.584}	& 0.405\\
		 \hline
		 CP UW 	 	& 0.881	& 0.892	& 0.900	& 0.885	& 0.844	& 0.023\\
		 CP W 	 		& \textbf{0.911}	& 0.875	& 0.899	& 0.889	& 0.770	& 0.023\\
		 \hline
		 OS UW 	 	& 0.697	& 0.672	& 0.665	& 0.671	& 0.652	& 0.397\\
		 OS W  		& 0.697	& 0.672	& 0.665	& 0.672	& 0.652	& \textbf{0.538}\\		
		 \hline \hline
	\end{tabular}
	\caption{Louvain method (LM), COPRA (CP) and OSLOM (OS) NMI performance for artificial networks with average degree $\langle k \rangle = 20$. Statistically significant improvements (\emph{p-value} $< 0.001$) are highlighted in bold.}
	\label{tab:artificial}
\end{table}

From the analysis of this table, we can draw the following conclusions:

\begin{enumerate}

\item If $\mu$ is low (i.e., $\mu = 0.1$ or $\mu = 0.2$), all the three approaches achieve a
    high NMI both with and without our weighting strategy. By contrast, for large values of
    $\mu$ (for example $\mu = 0.6$), the NMI deteriorates (especially for COPRA). In
    particular, if $\mu > 0.5$ a vertex has more neighbors outside the community to which it is
    assigned than in the community itself. Among the three methods, COPRA suffers the increase
    of $\mu$ more than Louvain method and OSLOM. In fact, if $\mu$ is around 0.1-0.2, the NMI
    achieved by COPRA is in line with that of Louvain method and in general it is quite high
    (between 0.849 and 0.917) but if $\mu$ tends to 0.6 its NMI decreases of about 90\% (and
    its values are around 0.018-0.023). This depends on the features of COPRA: in fact, a
    vertex $v$ is assigned to a community $C$ if most of its neighbors already belong to $C$.
    Of course, such an assignment is problematic for large values of $\mu$ because, as already
    observed, the neighbors of $v$ could be equally split across multiple communities.

\item It is worth observing that coupling Louvain method, COPRA and OSLOM with our strategy generally provides also an increase of NMI.
	To assess if the improvement provided by our method is significant or not, we carried out a t-test.
	We considered a \emph{p-value} lesser than 0.001 to determine, over 10 runs of each algorithm ($df = 9$), if the obtained increment was statistically significant or not. According to this test, significant increments are reported in bold in Table \ref{tab:artificial}.
	It emerges that the increase of NMI is generally extremely significant pairing the Louvain method with our weighting strategy (in particular, for $\gamma = 2$ this choice is able to guarantee significant improvements with low values of $\mu$, while for $\gamma = 3$ the increment is obtained also for high values of $\mu$.)
	Regarding the choice of COPRA paired with our strategy, we also obtain statistically significant improvements of NMI with all assignements of $\gamma$ and $\beta$, in particular with medium/high values of $\mu$.
	Differently, by considering OSLOM, the only extremely significant improvement in NMI is obtained while considering $\mu = 0.6$, but this increase is neat and appears in all possible configurations of $\gamma$ and $\beta$.

\item For a fixed value of $\gamma$ and $\mu$, we observe that the NMI achieved by the Louvain
    method decreases if $\beta$ ranges from 1 to 2. If $\beta$ gets larger, there are few
    communities containing a large number of vertices and a large number of communities which
    have roughly the same size because they contain few vertices. These communities are hard to
    find due to the so-called {\em resolution limit} \cite{fortunato2007resolution}: in
    particular, it is possible to show that community detection algorithms based on the
    principle of modularity maximization may fail to find communities containing less than
    $\sqrt{|E|/2}$ edges, being $|E|$ the number of edges in the entire network.

    However, a big result emerges from Table \ref{tab:artificial}: if we couple LM with the
    WERW-Kpath algorithm then the decay in NMI is softened. This depends on the different
    definition of the modularity function that the Louvain method attempts to optimize: in the
    case of weighted network, in fact, the term $A_{ij}$ is no longer 0 or 1 depending if an
    edge links the vertices $i$ and $j$ but it is a real number in [0,1] defining the strength
    of their links. The definition of the $Q$ function, therefore, is more precise and this
    allows higher values of $Q$ to be achieved.

    \end{enumerate}

\section{Related Works}
\label{sec:related}

In this section we describe some works related to our research.

First of all, we point out that an early version of the WERW-Kpath algorithm discussed appeared in \cite{ferrara2011novel}. We brought in some little modifications to the original WERW-KPath algorithm to achieve more solid results from a theoretical standpoint on the behavior of the algorithms itself.
In detail, in the algorithm presented in \cite{ferrara2011novel} the weight $\omega(e)$ of $e$ is proportional to the number of times $e$ is selected by the algorithm. This weight is interpreted as the edge centrality of $e$, i.e., we set $\hat{L}_k(e) = \omega(e)$: therefore, $\hat{L}_k(e)$ represents the {\em frequency} of selecting $e$ by means of random simple paths consisting of at most $\kappa$ edges. By contrast, in this paper, the weight $\omega(e)$ of $e$ counts how many times $e$ is selected and the edge centrality $\hat{L}_k(e)$ returned by the algorithm is equal to $\omega(e)$ divided by the number of trials $\rho$ performed by the algorithm. In this case, $\hat{L}_k(e)$ represents the {\em probability} of selecting $e$. This has relevant practical consequences. In fact, in \cite{ferrara2011novel}, we were able to prove that the edge $\omega(e)$ lies in a closed interval of the form $[\xi_1 L_{\kappa}(e), \xi_2 L_{\kappa}(e)]$, being $\xi_1$ and $\xi_2$ two real constants. However, we were not able to provide an estimation of $\xi_1$ and $\xi_2$, and, then, to quantify the approximation error associated with the estimation of $L_{\kappa}(e)$. In addition, we were not able to relate the number $\rho$ of trials carried out by the algorithm to the accuracy of the algorithm in approximating the actual edge centrality values. Some design considerations (later supported by experimental trials) suggested us to set $\rho = |E| - 1$.
Both these limitations have been addressed in the current version of the algorithm by means of Theorem \ref{th:bounds} and Corollary \ref{cor:bounds}. In fact, by means of Theorem \ref{th:bounds}, we provide tight bounds on the probability that $|\hat{L}_{\kappa}(e) - L_{\kappa}(e)|$ exceeds a given threshold and, by means of Corollary \ref{cor:bounds} and the subsequent reasoning we showed that, in case of real networks, a number of iterations equal to the number of vertices in the network was enough to provide accurate results.

However, the main novelty introduced in this paper is that we showed how to apply the WERW-Kpath algorithm to a non-trivial problem, i.e., the task of finding communities in networks. To the best of our knowledge, there are few works proposing to weight edges in a network to improve the quality of the community detection process.

One of the first approach to weighting edges was proposed in \cite{khadivi2010community}. In that
paper, the authors propose a modified version of the Girvan-Newman algorithm called {\em Newman
Fast}. In particular, given an unweighted, undirected graph $G$, each edge $e_{ij}$ connecting a
pair of vertices $i$ and $j$ is weighted. The weight of $e_{ij}$ is the normalized product of two
terms: the former is the inverse of the edge betweenness associated with $e_{ij}$ whereas the
latter (called {\em common neighbor ratio}) is the normalized number of vertices which are linked
to both $i$ and $j$. After that weighting step has been carried out, the Newman Fast algorithm
attempts to minimize the function $\tilde{Q}$ defined as

$$
\tilde{Q} = \sum_{l} (e_{ll} - a_l^2)
$$

The authors define $e_{lp}$ equal to half of the sum of the weights of the edges that start
from vertices in community $l$ and end in vertices located in community $p$ over the sum of the
weights of all the edges in the network. Therefore, $e_{ll}$ is the sum of the weights of the edges
contained within the community $l$. The parameter $a_l$ is defined as $a_l = \sum_k e_{lk}$. The
algorithm proceeds adopting a greedy strategy to maximize $\tilde{Q}$: initially, each vertex forms
a community and communities are merged so that to increase the value of $\tilde{Q}$. The process
stops when no further improvement of $\tilde{Q}$ can be achieved or all the vertices have been
inserted into a single community.

An improvement of the approach of \cite{khadivi2010community} was presented, by the same authors, in \cite{khadivi2011network}.
In that paper the authors suggested a slightly different weighting schema in which the contributions of edge betweenness and common neighbor ratio are combined through two weights $\alpha$ and $\beta$.
The authors suggest to tune $\alpha$ and $\beta$ so that to maximize the variance in edge distribution.

There are some differences between our approach and that of \cite{khadivi2010community}. In detail,
in the approach of \cite{khadivi2010community}, the edge weight is the product of two terms: the
former is the edge betweenness, which is a {\em global parameter} (i.e., its computation requires
to know the whole network topology) and the common neighbor ratio, which is a {\em local parameter}
(i.e., it can be computed by knowing only the neighbors of two vertices). Our $\kappa$-path
centrality, instead, lies between local and global measures because it can be computed by
considering random paths of length at most $\kappa$. Therefore, if $\kappa$ is kept low (resp.,
high) the edge centrality configure itself as a local (resp., global) measure. As a further
difference, the edge weighting procedure outlined in \cite{khadivi2010community,khadivi2011network}
has been used to design a modified version of $Q$ and a greedy algorithm to optimize it. By
contrast, in our approach we do not focus on any specific community detection algorithm and,
therefore, our approach can be used also with algorithms like COPRA or OSLOM which {\em do not
attempt} at maximizing modularity. In general, our strategy can be paired up with any community
detection algorithm handling weighted networks.

A further, interesting study, is presented in \cite{berry2011tolerating}. In that paper the authors
studied the aforementioned problem of resolution limit \cite{fortunato2007resolution}. In
\cite{fortunato2007resolution}, the authors showed that the size of the smallest community which
can be detected is $\sqrt{|E|/2}$, being $|E|$ the number of edges in the network. In
\cite{berry2011tolerating} the authors pointed out that by weighting a network the resolution limit
amounts to $\sqrt{W\varepsilon/2}$, being $W$ the sum of the weights in the network and
$\varepsilon$ the maximum weight of an edge connecting vertices located in two different
communities. Therefore, a wise choice of weights can significantly lower the resolution limit. The
authors suggested a weighting schema in which the weight of an edge depends on the number of cycles
of length $k$ containing that edge, being $k$ a fixed integer. The authors provides a modified
version of the Clauset, Newman and Moore algorithm \cite{clauset2004finding} capable of taking into
account edge weights.

Our approach differs from that proposed in \cite{berry2011tolerating}. In fact, we leverage on
random walks to compute edge weight whereas the approach of \cite{berry2011tolerating} relies on
the identification of cycles of length $k$. Unfortunately, the identification of these cycles can
be very time expensive as soon as $k$ gets large.
On the contrary, we proved both theoretically and experimentally that our strategy scales very well also if applied to large networks.

\section{Conclusions} \label{sec:conclusions}

In this paper we discussed an algorithm, called WERW-Kpath, to compute edge centralities in networks
and we showed that the strategy of weighting edges can generate a significant improvement in the
process of discovering communities. The WERW-Kpath algorithm exploits random walks of bounded
length to compute edge centralities. We provided a theoretical analysis of the behavior of the
WERW-Kpath algorithm and we showed how to use it in conjunction with already existing community
detection algorithm. We studied the merits and weaknesses of the WERW-Kpath algorithm by coupling
it with three state-of-the-art algorithms, namely Louvain Method, COPRA and OSLOM. Experiments carried out on real
networks show that 
our approach is able to improve the modularity of the community structure detected by the algorithms mentioned above and the improvement up to 17.3\%. We carried out also experiments on artificial networks: experiments showed that coupling Louvain method, COPRA and OSLOM with our strategy generally provides an increase of the Normalized Mutual Information.

As for future work, we plan to implement a multi-threaded version of the WERW-Kpath algorithm
so that we can simulate multiple random walks on the network in parallel. We plan to experimentally
study the computational improvements deriving from this choice. A further research direction
includes the creation of a friendship \emph{recommender system} which suggests new possible
connections to the users of a very large-scale online social network, based on the communities they
belong to.

\bibliographystyle{abbrv}


\end{document}